\documentclass[12pt]{iopart}
\usepackage{bm}
\usepackage{color}
\usepackage{psfig,epsfig}
\usepackage{cite}
\usepackage{subfigure}
\usepackage{amsmath-IOP}
\usepackage{amsfonts,amssymb,mathrsfs}

\begin{document}

\title{Inversion of the star transform}

\author{Fan Zhao}
\address{Department of Mathematics, University of Pennsylvania,
  Philadelphia, Pennsylvania 19104} 

\author{John C. Schotland} 
\address{Department of Mathematics and Department of Physics,
  University of Michigan, Ann Arbor, Michigan 48109}

\author{Vadim A. Markel}
\address{Departments of Radiology and Bioengineering and Graduate
  Group in Applied Mathematics and Computational Science, University
  of Pennsylvania, Philadelphia, Pennsylvania 19104}

\begin{abstract}
  We define the star transform as a generalization of the broken ray
  transform introduced by us in previous work. The advantages of using the
  star transform include the possibility to reconstruct the absorption
  and the scattering coefficients of the medium separately and
  simultaneously (from the same data) and the possibility to utilize
  scattered radiation which, in the case of the conventional X-ray
  tomography, is discarded. In this paper, we derive the star
  transform from physical principles, discuss its mathematical
  properties and analyze numerical stability of inversion. In
  particular, it is shown that stable inversion of the star transform
  can be obtained only for configurations involving odd number of
  rays. Several computationally-efficient inversion algorithms are
  derived and tested numerically.
\end{abstract}

\submitto{\IP}
\date{\today} 


\section{Introduction}
\label{sec:intro}

Image reconstruction techniques based on inversion of the Radon
transform are well-established. These methods generally neglect the
phenomenon of scattering and are based on the laws of geometrical
optics and Beer's law, which describes attenuation of rays upon
straight-line propagation. However, X-rays do experience significant
scattering in tissues and, at high energies, attenuation of X-rays is
predominantly explained by Compton scattering. Just like ballistic
photons, scattered photons carry information about the medium they
propagate through. Utilization of such photons can be advantageous.
In the case of X-rays, account of scattering is simplified because,
for the physical parameters encountered in typical applications, the
single-scattering approximation can be used safely. All this has
stimulated interest in using single-scattered photons for tomographic
imaging~\cite{norton_94_1,yuasa_97_1,wang_99_1,busono_99_1,cong_11_1,aviles_11_1}.
We also note that, under suitable conditions, single-scattering regime
can be applicable to optical imaging as well, e.g., in the mesoscopic
scattering regime~\cite{rossum_99_1}.

When scattering is taken into consideration, the photon
trajectories are no longer straight lines. Different curved or
piecewise linear trajectories have been explored, the type of
trajectory depending on energy- and angle-selectivity of sources and
detectors and on the choice of contrast mechanism. Thus, a series of
papers have explored a circular-arc transform, which arises when the
signal is generated by first-order Compton
scattering~\cite{norton_94_1,wang_99_1,driol_08_1,nguyen_10_1}. In the
imaging modality proposed in these references, the contrast mechanism
is related to the spatially-varying efficiency of Compton scattering
while attenuation of the scattered rays by the medium is neglected,
although an approximate correction to account for the attenuation can
be introduced~\cite{norton_94_1}. Mathematically, it was shown that
the generalized Radon transform on co-planar circles whose centers are
restricted to a bounded domain is invertible~\cite{agranovsky_96_1}.
Radon transforms on other smooth curves have also been
considered~\cite{maass_89_1,lissiano_97_1,maeland_98_1}.  In other
applications, the trajectories associated with detection of
single-scattered photons are piecewise
linear~\cite{yuasa_97_1,busono_99_1,cong_11_1,ilmavirta_13_1}. In
other cases, the arising transform involves area integrals rather than
integrals over well-defined trajectories~\cite{aviles_11_1}.

In this paper, we focus on an imaging modality introduced by us in
Refs.~\cite{florescu_09_1,florescu_10_1,florescu_11_1}, wherein
angularly-resolved sources and detectors are employed but no energy
resolution or sensitivity is assumed. The corresponding integral
transform of the medium is referred to as the broken ray Radon
transform or, in some cases, as the V-line
transform~\cite{morvidone_10_1,ambartsoumian_12_1}, since the photon
trajectories of interest resemble  the letter V. A single
broken ray consists of a vertex and two rays originating from the
vertex, which are translated without rotations when the vertex is
scanned. The imaging modality based on inverting the broken ray
transform does not require multiple projections in the traditional
sense. It turns out that it is sufficient to scan several
angularly-resolved sources and detectors on one or both sides of a
long strip. Another useful feature of this modality is the ability to
reconstruct the attenuation and the scattering coefficients of the
medium separately.  However, in the simplest geometry involving a
single broken ray (V-line) whose vertex is scanned over a
two-dimensional area~\cite{florescu_09_1,florescu_10_1,florescu_11_1},
the inverse problem is mildly ill-posed and this results in various
image artifacts.

Imaging methods that utilize broken rays have attracted considerable
recent
attention~\cite{ambartsoumian_12_1,katsevich_13_1,gouia-zarrad_14_1,haltmeier_14_1}.
An important result was obtained by Katsevich and
Krylov~\cite{katsevich_13_1} who have demonstrated that a linear
combination of several broken ray measurements can be used to derive a
purely local reconstruction algorithm that involves only first-order
derivatives of the data. In this paper, we also explore an approach
that utilizes linear combinations of broken ray measurements but take
a different approach to image reconstruction. Namely, we describe a
reconstruction algorithm in spatial Fourier domain where the inverse
solution depends on the data nonlocally. The motivation behind this
approach is three-fold. First, it allows one to use classical Tikhonov
regularization. Second, within the spatial Fourier method, measurements of
ballistic (nonscattered) rays can be easily combined with measurements of 
broken rays in the same image reconstruction algorithm.
Third, the method developed by us is more flexible with respect to choosing
the ray geometry.

In what follows, we show that the broken ray transform is a particular
case of the star transform (introduced in this paper),
which involves line integrals taken with different weights over
several rays originating from the same vertex. As in the case of a
single broken ray, the rays comprising the star are translated without
rotations when the vertex is scanned. The local method of Katsevich
and Krylov~\cite{katsevich_13_1} and other similar methods can be
obtained by taking linear combinations of the ray integrals with
vector coefficients (this involves a definition of a vector data
function). We will discuss some mathematical properties of the star
transform, derive several computationally efficient methods for its
inversion in spatial Fourier domain, analyze stability of the
inversion algorithms and illustrate the results with numerical
examples. In particular, we show that stable inversion of a scalar
star transform can be obtained only if the number of rays is odd. More
detailed stability conditions are also obtained.
 
The paper is organized as follows. The star transform is introduced in
Sec.~\ref{sec:geometry}. In Sec.~\ref{sec:derivation}, we explain how
the star transform can be related to the physical measurements taken
by angularly-resolved source-detector pairs. We also explain how the
star transform can be constructed to facilitate separate
reconstruction of the attenuation and the scattering coefficients.  In
Sec.~\ref{sec:katz}, we make a connection between the star transform
and the local reconstruction method of Katsevich and Krylov~\cite{katsevich_13_1}, and
describe a general framework for obtaining similar
methods. In Sec.~\ref{sec:Fourier}, we obtain the
Fourier-space representation of the star transform. 
In Sec.~\ref{sec:proj}, we explain
how the star transform can be combined with traditional projection
measurements without significant modifications of the reconstruction
algorithm. In Sec.~\ref{sec:methods}, we derive several fast
computational algorithms for inverting the star transform in Fourier
domain. The results of this section are, in fact, more general because
they apply to any matrix that is given by a sum of a term whose
inverse or pseudo-inverse is known and a finite number of separable
terms.  In Sec.~\ref{sec:stab} we analyze the numerical stability of inverting the star transform. We have obtained a few simple necessary conditions for
stability. One such condition is that the star transform must
contain an odd number of ray integrals. The results are illustrated
with numerical examples of Sec.~\ref{sec:sim}. Finally,
Sec.~\ref{sec:disc} contains a discussion and a summary.

\section{Imaging geometry and definition of the star transform}
\label{sec:geometry}

The imaging modality described in this paper reconstructs the
properties of a three-dimensional medium slice-by-slice, similarly to
the conventional methods of X-ray tomography (obviously, this analogy
does not apply to the helical Radon transform and similar
generalizations). In what follows, we assume that a slice $x = {\rm
  const}$ of a three-dimensional medium has been selected and will
focus on reconstructing the attenuation coefficient $\mu(y,z) =
\mu_s(y,z) + \mu_a(y,z)$ in that slice. Here $\mu_s$ and $\mu_a$ are
the scattering and absorption coefficients. Several angularly-resolved
sources and detectors are scanned along the lines $z=0$ and $z=L$. We
seek to reconstruct the function $\mu(y,z)$ between these two lines in
the open strip ${\mathbb S} = \{0 < z < L\}$. We will also show how
the scattering coefficient $\mu_s(y,z)$ can be separately recovered.

\begin{figure}
\centerline{
\subfigure[]{\epsfig{file=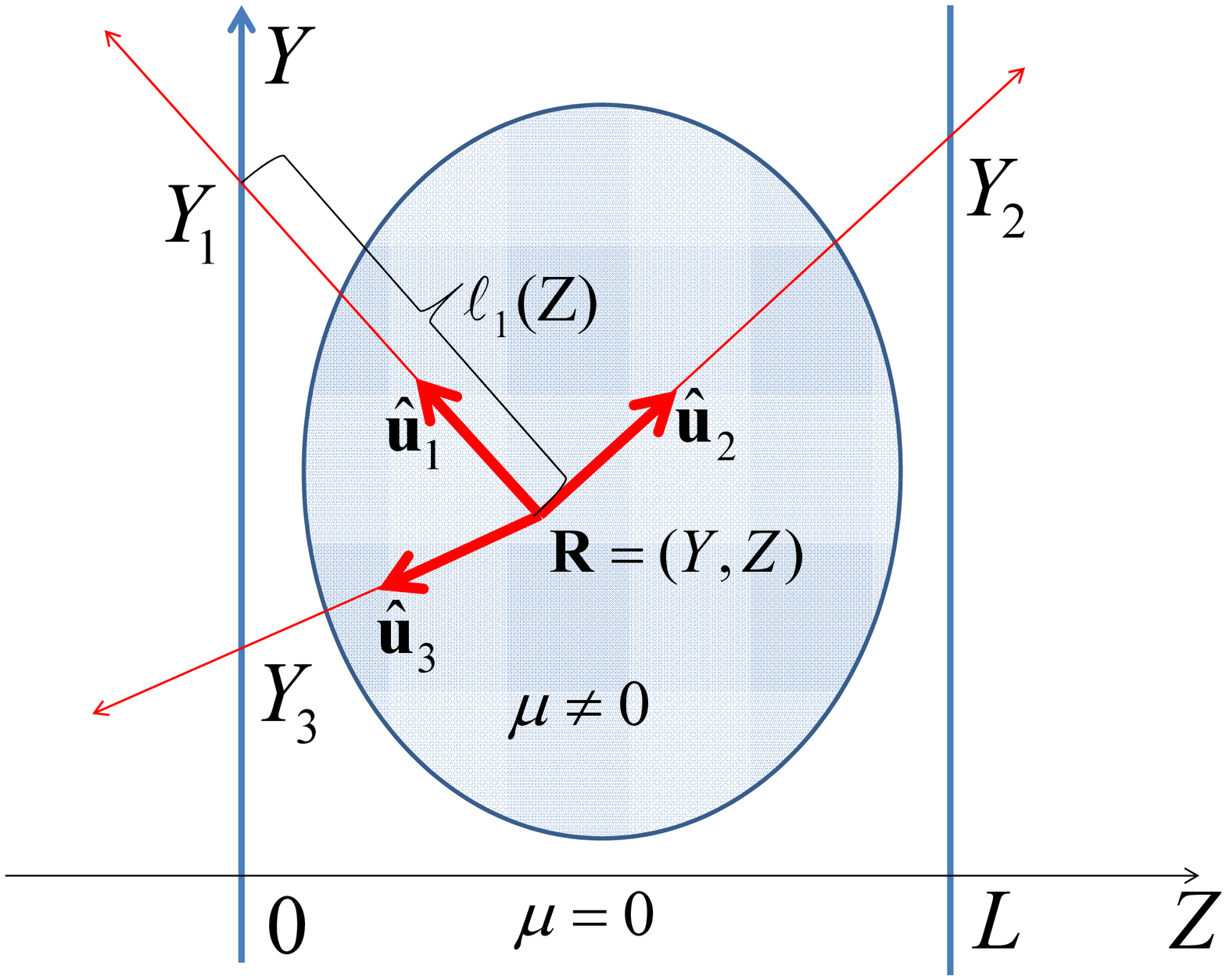,height=2.0in,clip=}}
\subfigure[]{\epsfig{file=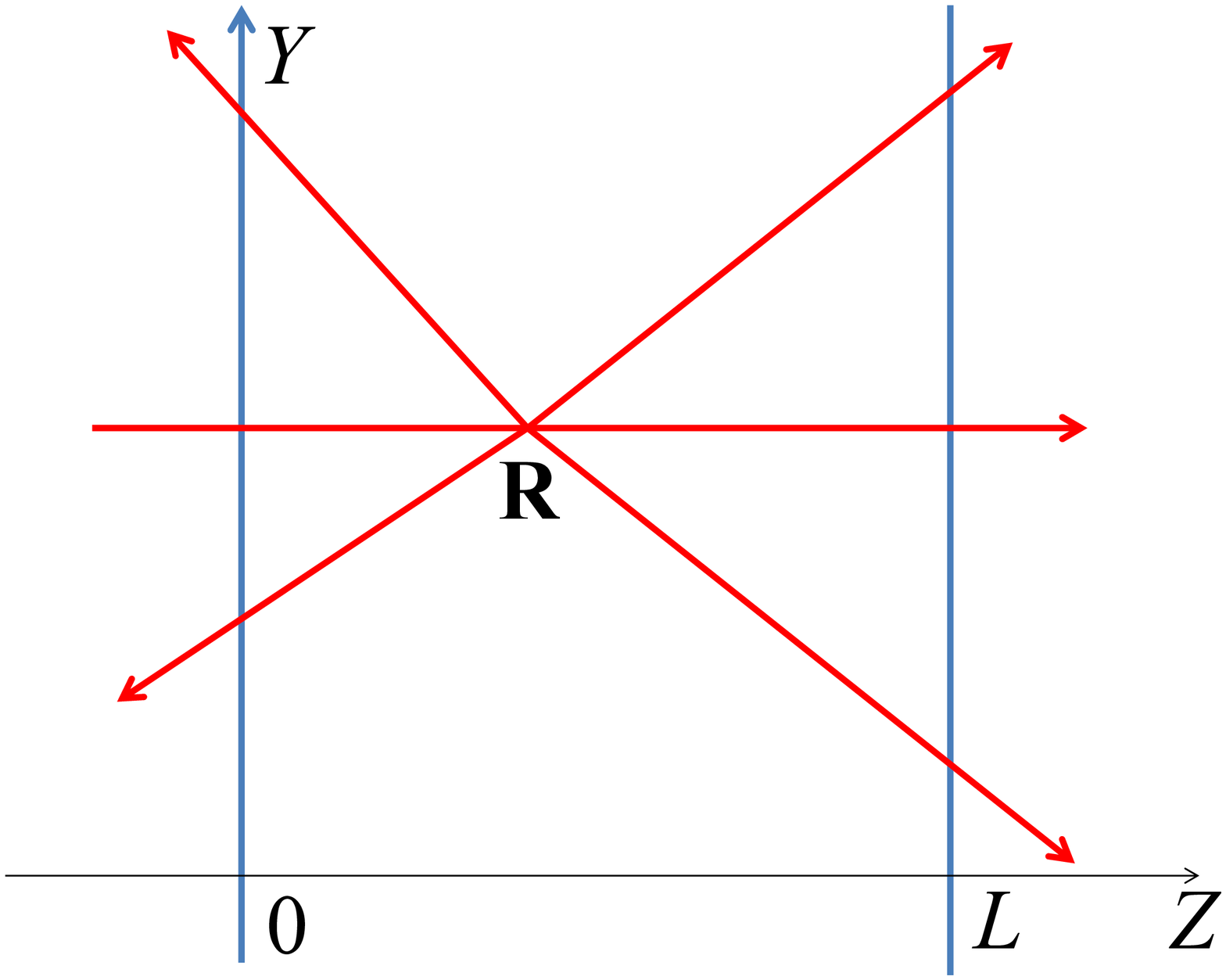,height=2.0in,clip=}}
}
\caption{\label{fig:geom} (color online) (a) Sketch of the imaging geometry for the case
  $N=3$ (the distances $\ell_2$ and $\ell_3$ are not shown).  (b)
  Imaging geometry in which simultaneous measurements of the ballistic
  and single-scattered rays (due to the same source) are utilized.}
\end{figure}

The transform whose inversion we study in this paper is of the
following form:
\begin{subequations}
\label{Star_Trans}
\begin{eqnarray}
\label{Phi_def}
\Phi({\bf R}) = \sum_{k=1}^K s_k I_k({\bf R}) \ , \ \ {\bf R}\equiv (Y,Z) \in
\bar{\mathbb S}  = \{0 \leq z \leq L\} \ \ ;  \\
\label{I_def}
I_k({\bf R}) = \int_0^{\ell_k(Z)} \mu \left({\bf R} + \hat{\bf u}_k
  \ell \right) d\ell \ . 
\end{eqnarray}
\end{subequations}
\noindent
Here $\Phi({\bf R})$ is the data function for a $K$-ray imaging
geometry, which is illustrated in Fig.~\ref{fig:geom}(a) for the
particular case $K=3$. It is defined in the closure of ${\mathbb S}$
(denoted here by $\bar{\mathbb S}$). Also, $\hat{\bf u}_k = (u_{ky},
u_{kz})$ is a set of $K$ unit vectors with nonzero projections onto
the $Z$-axis (that is, $u_{ky}^2 + u_{kz}^2 = 1$ and $u_{kz} \neq 0$);
$\ell_k(Z)$ is the distance (defined for each ray) from the vertex to
the boundary of ${\mathbb S}$; finally, $s_k \neq 0$ is a set of known
coefficients. The data function $\Phi({\bf R})$ is assumed to be known
(measured).

Since formation of broken rays depends on scattering, it is essential
that $\mu_s({\bf r}) > {\rm const} > 0$ for ${\bf r} \in \bar{\mathbb
  S}$ and this implies that $\mu({\bf r}) > {\rm const} > 0$ in
${\mathbb S}$. This is problematic for our purposes because $\mu({\bf
  r})$ does not have in this case a Fourier transform. Of course, the
scattering medium is never infinite in practice and one can introduce
physical boundaries to alleviate this problem. Then the data function
is no longer defined in an infinite strip. However, there exists a
simpler approach, which we will follow in this paper.  Namely, it is
physically reasonable to expect that the medium properties are
constant and known when $\vert y \vert \rightarrow \infty$. More
specifically, let
\begin{equation}
\label{mu_delta_mu}
\mu(y,z) = \bar{\mu} =
\bar{\mu}_s + \bar{\mu}_a \ \ \ \ {\rm for} \ \vert y \vert > y_{\rm
  max} > 0 \ ,
\end{equation}
\noindent
where bar is used to denote known background values of all
coefficients. This is true, in particular, if the inhomogeneities
$\mu({\bf r})$ are represented by a set of finite objects, as is often
the case in practice.  Then we can write $\mu({\bf r}) = \bar{\mu} +
\delta \mu({\bf r})$, where $\delta \mu({\bf r})$ is compactly
supported in a subset of ${\mathbb S}$.  This situation is
schematically illustrated in Fig.~\ref{fig:geom}(a). We emphasize that
the support of $\delta \mu$ or the value of $y_{\rm max}$ are not
known {\em a priori}. However, we can use (\ref{mu_delta_mu}) to
rewrite (\ref{Star_Trans}) as follows:
\begin{subequations}
\label{Star_Trans_delta}
\begin{align}
\label{Phi_def_del}
& \delta\Phi({\bf R}) \equiv \Phi({\bf R}) - \bar{\mu} \sum_{k=1}^K s_k
\ell_k(Z)  = \sum_{k=1}^K s_k \delta I_k({\bf R}) \ , \ \ {\bf R}\equiv (Y,Z)
\in {\mathbb S} \ \ ;  \\ 
\label{I_def_del}
& \delta I_k({\bf R}) = \int_0^{\ell_k(Z)} \delta \mu \left({\bf R} + \hat{\bf u}_k
  \ell \right) d\ell \ . 
\end{align}
\end{subequations}
\noindent
We notice that (\ref{Star_Trans_delta}) is exactly of the same form as
(\ref{Star_Trans}) except that some quantities have been redefined.
Since the functions $\ell_k(Z)$ are known, the left-hand side of
(\ref{Phi_def_del}) is also known and the expression for $\delta\Phi$
in terms of $\Phi$ and $\ell_k(Z)$ serves as a definition of the new
data function. In what follows, we will write for simplicity $\mu$
instead of $\delta\mu$, $\Phi$ instead of $\delta\Phi$, $I_k$ instead
of $\delta I_k$ and consider inversion of the transform
(\ref{Star_Trans}). However, we should keep in mind that, in this
formulation, $\mu$ is really the fluctuating part of the attenuation
coefficient and, therefore, it can be negative.  With these
definitions accepted, it is obvious that, as long as $\vert u_{kz}
\vert > {\rm const} >0$, we have $\Phi(Y,Z) = 0$ for $\vert Y \vert >
Y_{\rm max} > 0$. Note that $Y_{\rm max} \neq y_{\rm max}$, where the
latter quantity appears in (\ref{mu_delta_mu}), and the relation
between $Y_{\rm max}$ and $y_{\rm max}$ depends on the ray geometry.

Image reconstruction methods described in this paper allow one to
combine measurements of the type illustrated in
Fig.~\ref{fig:geom}(a), which employ single scattering at the vertex
${\bf R}$, with ballistic transmission measurements. This is
schematically illustrated in Fig.~\ref{fig:geom}(b). A family of
ballistic rays yields a single plane-parallel projection in a CT scan.
Since ballistic rays carry more photons than scattered rays,
measurements of the former are less affected by noise. Measurements of
ballistic rays that are perpendicular to the strip, as shown in
Fig.~\ref{fig:geom}(b), provide sufficient information to determine
the integral $\int_0^L \mu(y,z) dz$, which is related by Fourier
transform to the function $\mu_0(q)$ defined below in
Sec.~\ref{sec:Fourier}. Ballistic rays that enter the strip at the
angle $\theta$ to the normal provide information on the Fourier
coefficients of a more general form, namely, $\mu_n(\kappa_n {\rm
  ctg}\theta)$. Independent measurements of these coefficients can be
used in image reconstruction as is described in Sec.~\ref{sec:proj}.

\section{Physical principles and derivation of the star transform}
\label{sec:derivation}

We now explain what type of physical measurements are required to
obtain a transform of the type (\ref{Star_Trans}) and also discuss how
the scattering coefficient $\mu_s({\bf r})$ can be separately
recovered.

The mathematical concept of broken rays appears naturally when
transport of waves or particles through a medium is considered within
the first-order scattering approximation. Intensity carried by a
broken ray can be measured by source-detector pairs that are not on
axis. As is discussed in detail
in~\cite{florescu_09_1,florescu_10_1,florescu_11_1}, a broken ray is
defined when an angularly-resolved source and an angularly-resolved
detector are aligned so that the two rays drawn from the source and
the detector intersect in the slice of interest. The broken ray
consists of a vertex ${\bf R}=(Y,Z)$ and two rays originating from the
vertex and pointing to the source and the detector [see
Fig.~\ref{fig:geom}(a) for an illustration]. Each source and detector
is characterized by the directionality vector $\hat{\bf u}_k$ and each
can be scanned along the $Y$ axis. Consider a source-detector pair
characterized by the directionality vectors $\hat{\bf u}_j$ and
$\hat{\bf u}_k$. The projections of the source and detector positions
onto the $Y$ and $Z$ axes, $Y_j,Z_j$ and $Y_k,Z_k$, ($Z_j,Z_k=0,L$)
determine the broken ray geometry. Not all possible combinations of
$Y_j, Z_j, \hat{\bf u}_j; Y_k, Z_k, \hat{\bf u}_k$ have a vertex in
${\mathbb S}$. Source-detector arrangements without a vertex or with a
vertex outside of ${\mathbb S}$ do not generate useful data. If ${\bf
  R}\in {\mathbb S}$, then the power measured by the $j$-th detector
due to the $k$-th source, $W_{jk}$, is given by
\begin{equation}
\label{Int_BR}
W_{jk}({\bf R}) = W_0 S_{jk} \mu_s({\bf R}) \exp\left\{ - \left[ I_j({\bf R})
  + I_k({\bf R}) \right] \right\} \ , 
\end{equation} 
\noindent
where $W_0$ is the incident power generated by the source, which is
assumed to be the same for all sources, $I_j({\bf R})$ and $I_k({\bf
  R})$ are the integrals defined in (\ref{I_def}) and $S_{jk}$ is a
geometrical factor, which depends on $\cos \theta_{jk} = \hat{\bf u}_j
\cdot \hat{\bf u}_k$ but not on ${\bf R}$. The factors $S_{jk}$ can be
model-dependent, but we assume here that they are known.  For example,
if the power measured by the detector is described by the
single-energy radiative transport equation, an explicit expression for
$S_{jk}$ is given in~\cite{florescu_09_1}.

We now define data functions $\phi_{jk}({\bf R})$ according to
\begin{equation}
\label{phi_jk_def}
\phi_{jk}({\bf R}) = {\rm ln} \left[ \frac{W_{jk}({\bf R})}{W_0 S_{jk}
    \bar{\mu}_s} \right] \ , 
\end{equation}
\noindent
where $\bar{\mu}_s$ is the background scattering coefficient defined
in (\ref{mu_delta_mu}). It is important to note that, when a function
$\phi_{jk}({\bf R})$ is measured, one of the end-points of the
corresponding broken ray is a source and the other is a detector.
Physically, these are different devices. However, all data functions
$\phi_{jk}$ do not need to be acquired simultaneously. For example, in
the case $K=3$, the data functions $\phi_{12}$ and $\phi_{13}$ can be
acquired simultaneously using one source at $Y_1,Z_1$ and two
detectors at $Y_2,Z_2$ and $Y_3,Z_3$. Acquisition of the function
$\phi_{23}$ requires the use of a source at $Y_2,Z_2$ and a detector
at $Y_3,Z_3$ or {\em vice versa}, and can be performed separately.

Substituting (\ref{Int_BR}) into (\ref{phi_jk_def}), we obtain the
equation
\begin{equation}
\label{phi_ij_mu}
\phi_{jk}({\bf R}) = \left[ I_j({\bf R}) + I_k({\bf R}) + \eta({\bf R})
\right] (1 - \delta_{jk}) \ ,
\end{equation}
\noindent
where $\eta({\bf R}) = {\rm ln} \left[ \mu_s({\bf R}) / \bar{\mu}_s
\right]$. The diagonal terms $\phi_{kk}$ are not measurable; for
generality, we have defined these terms to be zero by including the
factor $(1 - \delta_{jk})$ in (\ref{phi_ij_mu}). Since $\phi_{jk} =
\phi_{kj}$, it is sufficient to consider only the pairs of indices
with $j < k$. If $K$ distinct directionality vectors are used, then
the number of independent functions $\phi_{jk}({\bf R})$ is
$K(K-1)/2$.

Consider first the case when the scattering coefficient is constant
and equal to $\bar{\mu}$, so that $\eta({\bf R}) = 0$. It is then easy
to see that
\begin{equation}
\label{sum_1}
\Phi \equiv \frac{1}{2(K-1)}\sum_{k=1}^K\sum_{j=1}^K \phi_{jk} =
\sum_{k=1}^K I_k \ .  
\end{equation}
\noindent
This equation is of the form (\ref{Phi_def}) with $s_k=1$. Therefore,
if transform (\ref{Phi_def}) defines the attenuation coefficient
uniquely, then the set $\phi_{jk}({\bf R})$ necessarily contains
redundant data points. Another way to obtain the star transform
(\ref{Phi_def}) with $s_k=1$ is to use cyclic summation. Let
$j(k)=k+1$ if $k<K$ and $j(k)=1$ if $k=K$. Then
\begin{equation}
\label{sum_2}
\Phi \equiv \frac{1}{2} \sum_{k=1}^{K} \phi_{k,j(k)} = \sum_{k=1}^K
I_k \ , 
\end{equation}
\noindent
which is of the same form as (\ref{sum_1}). We note that there are
many other linear combinations that result in the star transform with
$s_k \neq 1$.

We now turn to the case when $\mu_s({\bf R})$ can vary and,
correspondingly, $\eta({\bf R}) \neq 0$. Unlike in the case of uniform
scattering coefficient, the data functions $\phi_{jk}({\bf R})$
depends now on two unknown functions, $\mu({\bf R})$ and $\eta({\bf
  R})$. It is, in principle, possible to reconstruct both given a
sufficient number of degrees of freedom in the data. However, we wish
to simplify the inverse problem and exclude $\eta({\bf R})$ from the
equations analytically. In this case, introduction of the star
transform is a necessity rather than a choice. The approach
of~\cite{florescu_10_1,florescu_11_1,katsevich_13_1} was to make
linear combinations of the equations in (\ref{phi_ij_mu}) in such a
way as to exclude $\eta({\bf R})$; then, once the total attenuation
function is reconstructed, $\eta({\bf R})$ can be recovered from any
of the equations in (\ref{phi_ij_mu}). The linear combinations were
defined so that one of the ray integrals was excluded from the
resulting equations. For example, in a three-ray geometry, we have
defined~\cite{florescu_10_1,florescu_11_1} the data function as $\Phi
\equiv \phi_{12} - \phi_{13} = I_2 - I_3$. Thus, the scattering
contrast function $\eta({\bf R})$ was excluded but so was the ray
integral $I_1$. We note that the resultant two-ray geometry is {\em a
  priori} ill-posed, as is shown below in Sec.~\ref{subsec:q=Inf}.

Here we use a similar but somewhat more general
approach. We note that the problem of excluding $\eta$ while not
excluding any of the ray integrals from (\ref{phi_ij_mu}) can be
solved if we find a set of coefficients $c_{jk}$ such that (i)
$\sum_{jk} c_{jk} = 0$, (ii) $c_{kk}=0$, (iii) $c_{jk}=c_{kj}$, and
(iv) $s_k = \sum_j c_{jk} \neq 0$. Then
\begin{equation}
\label{sum_3}
\Phi \equiv \frac{1}{2}\sum_{j=1}^K \sum_{k=1}^K c_{jk} \phi_{jk} =
\sum_{k=1}^K s_k I_k  \ ,
\end{equation}
\noindent
where $\phi_{jk}$ satisfy (\ref{phi_ij_mu}). This is again a transform
of the form (\ref{Phi_def}) in which, by construction, $s_k \neq 0$.
Generally, there exist many different sets of $c_{jk}$ that satisfy
the above conditions. Some obvious coefficient examples for the cases
$K=3$ and $K=4$ are shown below:

\begin{equation*}
\begin{array}{rrr|r}
0 &  1 &  1 &  2 \\
1 &  0 & -2 & -1 \\
1 & -2 &  0 & -1 \\
\hline
2 & -1 & -1 & 0
\end{array} 
\hspace{2cm}
\begin{array}{rrrr|r}
0              &  1              & 1              & -1             & 1 \\
1              &  0              & -1             & -1             & -1 \\
1              &  -1             & 0              & 1              & 1 \\
-1             &  -1             & 1              & 0              & -1 \\
\hline
1 & -1 & 1 & -1 & 0
\end{array} 
\end{equation*}

\noindent
In the second example, all coefficients $c_{jk}$ and $s_k$ are equal
to $\pm 1$. Note that such arrangements are only possible if
$K(K-1)/2$ is even ($K=4,5,8,9,\ldots$).

\section{Star transform and the method of Katsevich and Krylov}
\label{sec:katz} 

If direct measurement of a single term $I_k({\bf R})$ was possible
(that is, if the data function with $K=1$ could be measured), then a
simple purely local image reconstruction algorithm involving one
first-order derivative could be obtained.  Indeed, we obviously have
\begin{equation}
\label{I_K1_diff_inv}
-(\hat{\bf u}_k \cdot \nabla_{\bf R}) I_k({\bf R}) = \mu({\bf R}) \ ,
\ \ {\bf R} \in {\mathbb S} \ . 
\end{equation}
Unfortunately, direct measurement of $I_k({\bf R})$ is physically
impossible. In the case of a spatially-uniform scattering coefficient
$\mu_{\rm s}({\bf R}) = {\rm const}$, we can define a star transform
by taking linear combinations of individual measurements in such a way
as to obtain an equation containing a single term of the form $I_k$.
For example, in the case $K=3$, we can take
\begin{equation}
\label{I3_exclusion}
\Phi \equiv \frac{1}{2}\left( \phi_{13} + \phi_{23} - \phi_{12} \right) = I_3 \ .
\end{equation}
\noindent
We can invert (\ref{I3_exclusion}) using the local formula
(\ref{I_K1_diff_inv}). Here the coefficients of the star transform are
$c_{13}=c_{23}=-c_{12}=1$.

In the more general and more practically-important case of
spatially-varying $\mu_s({\bf r})$, a single term $I_k({\bf R})$ can
not be mathematically ``isolated'' with the use of scalar coefficients
$c_{jk}$. However, we can define {\em vector} coefficients ${\bf
  c}_{jk}$ to obtain a local reconstruction algorithm involving only
first-order derivatives. In this case the data function
$\bm{\Phi}({\bf R})$ is also a vector and the reconstruction formula
is of the form 
\begin{equation}
\label{mu_local_Phi}
\mu({\bf R}) = - \dfrac{1}{\zeta}\nabla \cdot \bm{\Phi}({\bf R}) \ ,
\end{equation}
\noindent
where $\zeta$ is a coefficient. We will now outline a general approach
to obtaining such reconstruction formulas and provide a few examples.

Let a set of two-dimensional vector coefficients ${\bf c}_{jk}$
satisfy the following conditions: (i) $\sum_{jk} {\bf c}_{jk} = 0$,
(ii) ${\bf c}_{kk}=0$, (iii) ${\bf c}_{jk} = {\bf c}_{kj}$, and (iv)
${\bf s}_k = \sum_j {\bf c}_{jk} = \sigma_k \hat{\bf u}_k$, where
$\sigma_k$ is an arbitrary scalar. Here the conditions (i)-(iii) are
quite analogous to the similarly numbered conditions of the previous
section (for scalar coefficients) while condition (iv) is new: it
requires that ${\bf s}_k$ be collinear to the unit vector $\hat{\bf
  u}_k$. The conditions (i) and (iv) are consistent if $\sum_k
\sigma_k \hat{\bf u}_k = 0$. We can find the coefficients $\sigma_k$
that satisfy this condition if $K\geq 3$ (assuming all vectors
$\hat{\bf u}_k$ are different). Then define the vector data function
according to
\begin{equation}
\label{sum_4}
\bm{\Phi} \equiv \frac{1}{2}\sum_{j=1}^K \sum_{k=1}^K {\bf c}_{jk} \phi_{jk} \ ,
\end{equation}
\noindent
It can be seen that this data function satisfies the equation
\begin{equation}
\label{sum_5}
\bm{\Phi} = \sum_{k=1}^K \sigma_k \hat{\bf u}_k I_k  \ ,
\end{equation}
\noindent
which can be inverted according to (\ref{mu_local_Phi}) with $\zeta =
\sum_k \sigma_k$, assuming that the latter quantity is not zero.

As an example, consider the case $K=3$. The matrix of coefficients
that satisfy all the conditions stated above is
\begin{equation*}
\begin{array}{ccc|c}
0                       & \sigma_1 \hat{\bf u}_1 + \sigma_2 \hat{\bf u}_2 
                        & \sigma_1 \hat{\bf u}_1 + \sigma_3 \hat{\bf u}_3 
                        & \sigma_1 \hat{\bf u}_1  \\
\sigma_1 \hat{\bf u}_1 + \sigma_2 \hat{\bf u}_2 &  
0                       & \sigma_2 \hat{\bf u}_2 + \sigma_3 \hat{\bf u}_3
                        & \sigma_2 \hat{\bf u}_2 \\
\sigma_1 \hat{\bf u}_1 + \sigma_3 \hat{\bf u}_3 
                        & \sigma_2 \hat{\bf u}_2 + \sigma_3 \hat{\bf u}_3 &  0 
                        & \sigma_3 \hat{\bf u}_3 \\
\hline
\sigma_1 \hat{\bf u}_1  &
\sigma_2 \hat{\bf u}_2  &
\sigma_3 \hat{\bf u}_3  &
0
\end{array} 
\end{equation*}
\noindent
where the coefficients $\sigma_k$ are assumed to satisfy $\sigma_1
\hat{\bf u}_1 + \sigma_2 \hat{\bf u}_2 + \sigma_3 \hat{\bf u}_3 = 0$.
Application of this scheme results in the equation
\begin{eqnarray}
\label{local_rec_1}
\mu &= - \dfrac{1}{\zeta} \nabla \cdot \left[
\sigma_1 \hat{\bf u}_1 (\phi_{12} + \phi_{13}) + \sigma_2 \hat{\bf u}_2
(\phi_{12} + \phi_{23}) + \sigma_3 \hat{\bf u}_3 (\phi_{13} +
\phi_{23}) \right] \nonumber \\
    &= \dfrac{1}{\zeta} \nabla \cdot \left[ 
    \sigma_1\hat{\bf u}_1 (\phi_{23} - \phi_{12}) + 
    \sigma_2\hat{\bf u}_1 (\phi_{12} - \phi_{13})
    \right] \ .
\end{eqnarray}
\noindent
In particular, if $\hat{\bf u}_1 + \hat{\bf u}_2 + \hat{\bf u}_3 = 0$,
we can use $\sigma_1 = \sigma_2 = \sigma_3 = 1$ and the above equation
takes the simple form
\begin{align*}
\mu = \frac{1}{3} \nabla \cdot \left[ \hat{\bf u}_1 (\phi_{23} -
  \phi_{12}) + \hat{\bf u}_2 
(\phi_{13} - \phi_{12}) \right] \ .
\end{align*}

The method of Katsevich and Krylov~\cite{katsevich_13_1} can also be
derived using the mathematical formalism of this section. The
particular implementation of Ref.~\cite{katsevich_13_1} utilizes four
rays, $K=4$, but one of the scalar coefficients, say $\sigma_1$, is
zero, so that the corresponding ray does not enter the star transform.
The coefficient matrix utilized by Katsevich and Krylov is
\begin{equation}
\begin{array}{rrrr|r}
0              &  \sigma_2 \hat{\bf u}_2 & \sigma_3 \hat{\bf u}_3 &
\sigma_4 \hat{\bf u}_4 &  0              \\
\sigma_2 \hat{\bf u}_2 &  0              & 0              & 0
& \sigma_2 \hat{\bf u}_2 \\
\sigma_3 \hat{\bf u}_3 &  0              & 0              & 0
& \sigma_3 \hat{\bf u}_3 \\
\sigma_4 \hat{\bf u}_4 &  0              & 0              & 0
& \sigma_4 \hat{\bf u}_4 \\
\hline
0              & \sigma_2 \hat{\bf u}_2  & \sigma_3 \hat{\bf u}_3 &
\sigma_4 \hat{\bf u}_4 & 0
\end{array} 
\end{equation}
\noindent
where $\sigma_1 = 0$ and the remainder of the coefficients satisfy
$\sigma_2 \hat{\bf u}_2 + \sigma_3 \hat{\bf u}_3 + \sigma_4 \hat{\bf
  u}_4 = 0$. The corresponding inversion formula is
\begin{eqnarray}
\label{local_rec_2}
\mu &= - \dfrac{1}{\zeta} \nabla \cdot 
\left(
\sigma_2 \hat{\bf u}_2 \phi_{12} + \sigma_3 \hat{\bf u}_3 \phi_{13} +
\sigma_4 \hat{\bf u}_4 \phi_{14} \right) \nonumber \\
    &=\dfrac{1}{\zeta} \nabla \cdot 
    \left[
    \sigma_2 \hat{\bf u}_2 (\phi_{14}-\phi_{12}) +
    \sigma_3 \hat{\bf u}_3 (\phi_{14}-\phi_{13})
    \right]
\ .
\end{eqnarray}
\noindent
The difference between formulas (\ref{local_rec_1}) and
(\ref{local_rec_2}) is that these formulas utilize different
``individual measurements'' $\phi_{jk}$ and, therefore, a different
physical arrangement of sources and detectors. 

\section{Fourier basis representation}
\label{sec:Fourier}

We define the Fourier transform of $\mu(y,z)$ as follows:
\begin{subequations}
\label{FT_def}
\begin{align}
\label{ft_mu_inv}
& \mu(y,z) = \int_{-\infty}^\infty \frac{dq}{2\pi} e^{iqy} \tilde{\mu}(q,z) 
= \int_{-\infty}^{\infty} \frac{dq}{2\pi} e^{iqy }
 \frac{1}{L} \sum_{n=-\infty}^\infty \mu_n(q) e^{i \kappa_n z} \ , \\
\label{ft_mu_dir}
& \mu_n(q) = \int_{-\infty}^\infty dy e^{-iqy} \int_0^L dz 
e^{-i \kappa_n z} \mu(y,z) \ , \\
& \kappa_n = \frac{2\pi n}{L} \ .
\end{align}
\end{subequations}
\noindent
The same convention can be used for the data function $\Phi(Y,Z)$. We
assume that all functions are sufficiently ``nice'' so that the
Fourier transforms exist. It should be noted that, if
(\ref{Star_Trans}) is defined on an infinite plane, then the Fourier
transform of $\Phi(Y,Z)$ does not exist even if we include generalized
functions into consideration. When (\ref{Star_Trans}) is defined in
${\mathbb S}$ and all rays cross the strip boundaries, this problem is
removed. However, we should keep in mind that $\Phi(Y,0) \neq
\Phi(Y,L)$. We define
\begin{equation}
\label{Delta_def}
\Delta(Y) \equiv \frac{1}{2}\left[\Phi(Y,0) + \Phi(Y,L) \right] \ ,
\end{equation}
\noindent
Then
\begin{equation}
\label{Sum_Phi_n}
\sum_{n=-\infty}^\infty \Phi_n(q) = L \tilde{\Delta}(q) = L
\int_{-\infty}^\infty \Delta(Y) e^{-iqY} dY \ .
\end{equation}

We start by Fourier-transforming (\ref{Star_Trans}) along the
$Y$-direction:
\begin{align}
\tilde{\Phi}(q,Z)  
& = \sum_{k=1}^K s_k \int_0^{\ell_k(Z)} d\ell \int_{-\infty}^\infty
\mu(Y+u_{ky}\ell, Z+u_{kz}\ell) e^{-iqY} dY \nonumber \\
& = \sum_{k=1}^K s_k \int_0^{\ell_k(Z)} e^{iq u_{ky} \ell}
    \tilde{\mu}(q, Z+u_{kz}\ell) d\ell \nonumber \\ 
& = \sum_{k=1}^K \frac{s_k}{u_{kz}} e^{-i\beta_k(q) Z} \int_Z^{\xi_k}
e^{i\beta_k(q) z} \tilde{\mu}(q,z) dz \nonumber \\
& = \sum_{k=1}^K \frac{s_k}{u_{kz}} e^{-i\beta_k(q) Z}
\sum_{n=-\infty}^\infty \frac{e^{i[\beta_k(q)+\kappa_n]\xi_k} -
  e^{i[\beta_k(q)+\kappa_n]Z}}{i[\beta_k(q) + \kappa_n]} \mu_n(q) \ .
\label{Phi_FT}
\end{align}
\noindent
In the above derivation, we have taken advantage of the fact that the
upper limit of integration over $\ell$, $\ell_k(Z)$, is independent of
$Y$ and introduced the notations
\begin{equation}
\label{beta_xi_def}
\beta_k(q) = q \frac{u_{ky}}{u_{kz}} \ , \ \ \ \xi_k = \left\{ \begin{array}{ll}
L \ , &  {\rm if} \ u_{kz} > 0 \\
0 \ , &  {\rm if} \ u_{kz} < 0 \\
\end{array}\right. \ .
\end{equation}
\noindent
Here $\xi_k$ is the $Z$-coordinate of the $k$-th ray intersection with
the boundary of ${\mathbb S}$. Since we assume that all rays intersect
the strip boundaries, the quantities in (\ref{beta_xi_def}) are well
defined. It is useful to keep in mind that $\exp(i \xi_k \kappa_n) =
1$ for all values of indices.

Equation (\ref{Phi_FT}) is parameterized by $q$. To shorten the
notations, we will omit the explicit dependence on $q$ below by
writing $\beta_k$, $\mu_n$ instead of $\beta_k(q)$, $\mu_n(q)$, etc.,
except in a few special cases. We then fix $q$ and take the Fourier
transform of (\ref{Phi_FT}) with respect to $Z$, which results in the
following infinite system of linear equations:
\begin{equation} 
\label{Phi_m_mu_n}
\Phi_n = \mu_n \sum_{k=1}^K \frac{i s_k}{u_{kz}(\beta_k + \kappa_n)}
+ \sum_{k=1}^K \frac{s_k e^{i\beta \xi_k} \left(e^{-i\beta_k L } -
    1\right) }{L u_{kz} (\beta_k + \kappa_n)} \sum_{m=-\infty}^\infty
\frac{\mu_m} {\beta_k + \kappa_m}  \ .
\end{equation}
\noindent
Introducing the notations
\begin{equation}
\label{d_alpha_def}
d_n = \sum_{k=1}^K \frac{i s_k}{u_{kz}(\beta_k + \kappa_n)} =
\sum_{k=1}^K \frac{i s_k}{\hat{\bf u}_k \cdot (q,\kappa_n)} \ , \ \
\alpha_k = \frac{e^{i\beta_k\xi_k}\left( e^{-i\beta_k
    L} - 1 \right)}{L u_{kz}}  \ ,
\end{equation}
\noindent
we can rewrite (\ref{Phi_m_mu_n}) in the form
\begin{equation} 
\label{Phi_m_mu_n_1}
\Phi_n = d_n \mu_n 
+ \sum_{k=1}^K \frac{s_k \alpha_k}{\beta_k + \kappa_n} \sum_{m=-\infty}^\infty
\frac{\mu_m} {\beta_k + \kappa_m}  \ .
\end{equation}
\noindent
In (\ref{d_alpha_def}), $(q,\kappa_n)$ is the two-dimensional Fourier
vector. It can be seen that (\ref{Phi_m_mu_n_1}) is an infinite set of
algebraic equations whose matrix is a sum of one diagonal matrix and
$K$ separable matrices. It will also prove useful to introduce Dirac
notations. Let
\begin{subequations}
\begin{eqnarray}
\label{mu_vect_def}
& \vert \mu \rangle = \left( \ldots,  \mu_{-1},  \mu_0,  \mu_1, \ldots
 \right)^T \ , \\
\label{Phi_vect_def}
& \vert \Phi \rangle = \left( \ldots,  \Phi_{-1},  \Phi_0,  \Phi_1, \ldots
 \right)^T \ , \\
\label{a_vect_def}
& \vert a_k \rangle = \left(\ldots, \frac{1}{\beta_k + \kappa_{-1}},
  \frac{1}{\beta_k  + \kappa_0}, \frac{1}{\beta_k + \kappa_1},
  \ldots \right)^T \ .
\end{eqnarray}
\end{subequations}
\noindent
Then (\ref{Phi_m_mu_n_1}) can be written as
\begin{equation} 
\label{main}
\vert \Phi \rangle = D \vert \mu \rangle
+ \sum_{k=1}^K s_k \alpha_k \vert a_k \rangle \langle a_k \vert
\mu \rangle \equiv A\vert \mu \rangle \ .
\end{equation}
\noindent
Here the diagonal matrix $D$ has the elements $D_{nm} = d_n
\delta_{nm}$ and the second equality defines the matrix $A$.

\section{Star transform combined with projection measurements}
\label{sec:proj}

The function $\mu_0(q)$ can be measured with relatively high precision
by utilizing ballistic (non-scattered) rays as is illustrated in
Fig.~\ref{fig:geom}(b).  Measurements of ballistic rays and of the
single-scattered rays can be performed simultaneously, without the
need to employ additional sources of radiation. Therefore, information
on $\mu_0(q)$ can be obtained as long as at least one of the sources
is oriented perpendicularly to the strip, as shown in the figure. One
motivation for developing the imaging modality described here is to
reduce the radiation dose received by a patient by reducing the number
of projections employed. In this respect, discarding ballistic photons
is not an efficient approach. Mathematically, the knowledge of
$\mu_0(q)$ does not determine the function $\mu(y,z)$ [to this end,
the whole set of coefficients $\mu_n(q)$ is required], but it does
improve the conditioning of the inverse problem. Therefore, it is
reasonable to assume that the projection measurements illustrated in
Fig.~\ref{fig:geom}(b) have been performed and that $\mu_0(q)$ is
known. Then we can rewrite (\ref{Phi_m_mu_n_1}) as follows:
\begin{subequations}
\label{Psi_m_mu_n_1}
\begin{align} 
\label{Psi_0}
& \Psi_0 = \sum_{k=1}^K \frac{s_k \alpha_k}{\beta_k} \sum_{m\neq 0}
\frac{\mu_m}{\beta_k + \kappa_m} \ , &n=0 \ , \\
\label{Psi_n}
& \Psi_n = d_n \mu_n 
+ \sum_{k=1}^K \frac{s_k \alpha_k}{\beta_k + \kappa_n} \sum_{m\neq 0}
\frac{\mu_m} {\beta_k + \kappa_m}  \ , & n\neq 0 \ ,
\end{align} 
\end{subequations}
\noindent
where
\begin{subequations}
\label{Psi_def}
\begin{align}
& \Psi_0 \equiv \Phi_0 - \mu_0 \left( d_0 + \sum_{k=1}^K 
\frac{s_k \alpha_k}{\beta_k^2} \right) \ , & n=0 \ , \\
& \Psi_n \equiv \Phi_n - \mu_0 \sum_{k=1}^K 
\frac{s_k \alpha_k}{\beta_k(\beta_k + \kappa_n)} \ , & n \neq 0 \ .
\end{align}
\end{subequations}
\noindent
The right-hand side of (\ref{Psi_m_mu_n_1}) depends only on $\mu_n$
with $n\neq 0$; therefore, this set (when truncated so that $\vert n
\vert \leq n_{\rm max}$; see more detail on truncation in
Sec.~\ref{subsec:trunc}) has more equations than unknowns. One
possible approach is to disregard equation (\ref{Psi_0}). In an ideal
setting, e.g., if the data are generated using inverse crime and
contain no noise or systematic errors, all equations in
(\ref{Psi_m_mu_n_1}) must be consistent and the disregard of
(\ref{Psi_0}) does not affect the solution. However, if experimental
measurements are used, then (\ref{Psi_0}) provides an additional
useful constraint and it might be advantageous to seek the
pseudo-inverse of the overdetermined system, as is described in
Sec.~\ref{subsec:A_pseudo} below.

We finally note that measurement of ballistic rays that enter the slab
at the angle $\theta$ can also be utilized in similar manner. Such
rays yield independent measurements of the Fourier coefficients
$\mu_n(q=\kappa_n {\rm ctg}\theta)$.

\section{Methods of solution}
\label{sec:methods}

We now discuss two efficient algorithms for solving equations of the
type (\ref{main}). However, we will consider in this section inversion
algorithms for a more general matrix $A$, which is of the form
\begin{equation}
\label{A_gen}
A = D + V \ , \ \  V = \sum_{k=1}^K \vert b_k \rangle \langle  a_k \vert \ ,
\end{equation}
where $\vert a_k \rangle$ and $\vert b_k \rangle$ are not necessarily
collinear. The special case of the star transform is obtained if we
take $\vert b_k \rangle = s_k \alpha_k \vert a_k \rangle$ and assume
that $D$ is the diagonal matrix defined by (\ref{d_alpha_def}). The
size of $A$ can be either finite or infinite. In particular, $A$ can
be a finite-size rectangular or square matrix. In the case of star
transform, these details depend on the type of truncation of the
infinite set of equations (\ref{main}). Truncation is discussed in
Sec.~\ref{subsec:trunc}. In Sec.~\ref{subsec:A_pseudo}, we describe an
algorithm for fast iterative computation of the Tikhonov-regularized
pseudo-inverse of a finite-size rectangular $A$. An interesting
feature of this algorithm is that it does not require the knowledge or
computation of the singular vectors and singular values of $A$.
Finally, in Sec.~\ref{subsec:dir} we describe a method for direct
inversion of $A$. In this method, $A$ can be either finite and square
or infinite. The power of this method is that it allows to invert an
infinite-dimensional matrix, essentially, without any approximations.
However, it is not possible to measure all components of the
infinite-dimensional vector of data $\vert \Phi \rangle$ and the
components that are not measurable are approximated by zeros.

\subsection{Truncation}
\label{subsec:trunc}

Here we discuss various ways to truncate the infinite set of equations
(\ref{main}).

To perform image reconstruction on a finite grid, it is sufficient to
know the coefficients $\mu_n$ for a finite range of $n$.  For example,
if the image is discretized in the $Z$-direction using $2 n_{\rm
  max}+1$ points $z_n=h n$, $n=0,1,\ldots, 2 n_{\rm max}$,
$h=L/2n_{\rm max}$, then we only need to know $\mu_n$ for $-n_{\rm
  max} \leq n \leq n_{\rm max}$. If we sample the data function on the
same grid, we have access to the quantities $\Phi_n$ with $n$ in the
same range. Other components of $\vert \Phi \rangle$ are in this case
not measurable. If the matrix $A$ in (\ref{main}) were block-diagonal,
with one block encompassing all indexes $n$ that satisfy the above
inequality, then we could have used the available measurements of
$\Phi_n$ to reconstruct the required $\mu_n$ without any
approximations. In this case, truncation of (\ref{main}) would have
been trivial.

However, $A$ is not block-diagonal. As a result, the coefficients
$\mu_n$ with $-n_{\rm max} \leq n \leq n_{\rm max}$ depend on {\em
  all} measurements $\Phi_n$. Therefore, there exist two different
methods to truncate (\ref{main}). The first method is to set $\mu_n =
0$ for $\vert n \vert > n_{\rm max}$ and to disregard all equations
with $\vert n \vert > n_{\rm max}$. In this approach, {\em all}
equations in (\ref{main}) are changed, not only the ones that have
been disregarded. The second approach is to keep all equations in
(\ref{main}) but to substitute the unavailable measurements
$\Phi_n(q)$ with zeros. This second approach does not modify the
matrix $A$ but it makes an approximation of the data.

Normally, this second approach would not be practically feasible
because infinite sets of equations can not be handled numerically.
For the case at hand, however, we can utilize the known algebraic
structure of $A$, as is described in Sec.~\ref{subsec:dir} below.

\subsection{Iterative computation of the pseudo-inverse}
\label{subsec:A_pseudo}

Here we consider the case when $A$ is finite and either square and
singular or non-square. In both cases, inverse of $A$ may not exist.
Non-square $A$ is encountered, for example, if projection measurements
are used to compute some of the coefficients $\mu_n(q)$, as is
described in Sec.~\ref{sec:proj}. We therefore wish to derive an
inversion algorithm that is suitably regularized and contains no
numerical instabilities. We note that none of the terms in the
right-hand side of (\ref{A_gen}) can be expected to commute and,
therefore, it is not possible to find analytically the singular value
decomposition of $A$ even in the simplest case of one separable term.
The proposed method computes the pseudo-inverse of $A$ rather than its
singular-value decomposition, the latter being a more computationally
demanding yet an unnecessary task.

The main idea of this iterative computation of the pseudo-inverse is
the following.  Let us define the ``forward'' recursion as the set of
equations
\begin{equation}
\label{direct_recursion}
D_k = D_{k-1} + \vert b_k \rangle \langle a_k \vert \ , \ \ k=1,2,
\ldots , K \ ,
\end{equation}
\noindent
where $D_0 = D$. It can be seen that $D_K = A$, where $A$ is defined
by (\ref{A_gen}) and contains $K$ separable terms. Now let us assume
that we know the pseudo-inverse of $D_{k-1}$, denoted here by 
$D_{k-1}^+$. Then we can
compute the pseudo-inverse of the right-hand side of
(\ref{direct_recursion}) analytically using the formulas given below.
This will yield an expression for the pseudo-inverse of $D_k$ in terms
of the pseudo-inverse of $D_{k-1}$. This rule for pseudo-inverses can
be referred to as the ``inverse'' recursion. The inverse recursion can
be started easily because the pseudo-inverse of $D_0=D$ (a diagonal
matrix) is known. Then the inverse recursion can be followed for $K$
steps to compute $A^{+} = D_K^{+}$.

Let $A$ and all $D_k$ be finite $N \times M$ matrices. The
Tikhonov-regularized pseudo-inverse of $D_k$ can be defined as
follows:
\begin{equation}
\label{A_def}
D_k^+ = D_k^* S_{N,k} = S_{M,k} D_k^* \ ,
\end{equation}
\noindent
where
\begin{equation}
\label{S_k_def}
 S_{N,k} = (D_k D_k^* + \lambda^2 {\mathbb I}_N)^{-1} \ , \ \ S_{M,k}
 = (D_k^* D_k + \lambda^2 {\mathbb I}_M)^{-1} \ ,  
\end{equation}
\noindent
Here ${\mathbb I}_N$ and ${\mathbb I}_M$ are the identity matrices of
the size $N \times N$ and $M\times M$, respectively, and $\lambda$ is
the regularization parameter. We note that the inverses in
(\ref{S_k_def}) exist in the usual sense as long as $\lambda>0$.

The recursion starts with computing $S_{N,0}$ and $S_{M,0}$ according
to (\ref{S_k_def}), where $D_0 = D$. In the case of star transform,
$S_{N,0}$ and $S_{M,0}$ are diagonal matrices with the elements
\begin{equation}
\label{s_m_def}
s_m = \frac{1} {\vert d_m \vert^2 + \lambda^2} \ . 
\end{equation}
\noindent
In the case $N>M$ (an overdetermined problem), the matrix $S_{N,0}$
has $s_m$ in the first $M$ diagonal positions and $1/\lambda^2$ in the
positions $m=M+1,\ldots,N$, while $S_{M,0}$ is the $M\times M$ minor
of $S_{N,0}$.

The generic inverse iteration step requires a formula for updating
$S_{N,k+1}$, $S_{M,k+1}$ and $D_{k+1}^+$ in terms of $S_{N,k}$,
$S_{M,k}$, and $D_k^+$. This recursive rule is of the form
\begin{subequations}
\label{S_recursion}
\begin{align}
& S_{N,k+1} = S_{N,k} - S_{N,k} T_{N,k} S_{N,k} \ , \\ 
& S_{M,k+1} = S_{M,k} - S_{M,k} T_{M,k} S_{M,k} \ , \\ 
& D_{k+1}^+ = D_{k+1}^*S_{N,k+1} = S_{M,k+1} D_{k+1}^* \ .
\end{align}
\end{subequations}
\noindent
In the last formula, any of the two equivalent expressions for
$D_{k+1}^+$ can be used. Note that we have encountered two distinct
(but related by a permutation of notations) matrices $T_{N,k}$ and
$T_{M,k}$, which can be referred to as the pseudo-T matrices. Unlike
the true T-matrix, the pseudo-T matrices are guaranteed to exist as
long as $\lambda>0$. Analytical expressions for the pseudo-T matrices
can be obtained by tedious but straightforward algebraic calculation.
Here we adduce the final result:
\begin{subequations}
\label{main_iter}
\begin{align}
T_{M,k} = & \frac{1}{{\mathscr D}_k} \Big{[} 
\gamma_k \vert a_{k+1} \rangle \langle b_{k+1} \vert  D_k
 + {\rm H.c.} \nonumber \\
& + \lambda^2 P_k  \vert a_{k+1} \rangle \langle a_{k+1} \vert
  - Q_k D_k^* \vert b_{k+1} \rangle \langle b_{k+1} \vert D_k \Big{]}
\ , \\ 
T_{N,k} = & \frac{1}{{\mathscr D}_k} \Big{[}
\gamma_k D_k \vert a_{k+1} \rangle \langle b_{k+1} \vert
 + {\rm H.c.} \nonumber \\
&   + \lambda^2 Q_k \vert b_{k+1} \rangle \langle b_{k+1}\vert - P_k 
D_k \vert a_{k+1} \rangle \langle a_{k+1} \vert D_k^* \Big{]} \ ,
\end{align}
\end{subequations}
\noindent
where ''H.c.'' stands for Hermitian conjugate of the preceding term and
\begin{subequations}
\begin{align}
  & \gamma_k  = 1 +  \langle a_{k+1} \vert D_k^+ \vert b_{k+1} \rangle \ , \\
  & P_k = \langle b_{k+1} \vert S_{N,k} \vert b_{k+1} \rangle >0 \ , \
  \ Q_k = \langle a_{k+1} 
  \vert S_{M,k} \vert a_{k+1} \rangle > 0 \ , \\
  & {\mathscr D}_k = \vert \gamma_k \vert^2 + \lambda^2 P_k Q_k > 0 \ .
\end{align}
\end{subequations}
\noindent
As can be seen, the quantities numbers $P_k$, $Q_k$ and ${\mathscr
  D}_k$ are guaranteed to be positive. Correspondingly, the main
iteration step (\ref{S_recursion}) is always well-defined. The
iterations are formally terminated at $k=K$ and the final result is
obtained as $A^+ = D_K^+$.

Note that the algorithm described here involves numerical operations
on matrices and vectors whose elements are not known analytically.
Therefore, this algorithm can not be used in the infinite-dimensional
case.

\subsection{Solution by direct matrix inversion}
\label{subsec:dir}

Here we assume that $A$ is of infinite size, although the algorithm is
also applicable if $A$ is finite and square.

We wish to solve the equation $A\vert \mu \rangle = \vert \Phi
\rangle$, where $A$ is of the form (\ref{A_gen}). Let $x_k = \langle
a_k \vert \mu \rangle$, $k=1,\ldots,K$. Then multiply the equation by
$\langle a_j \vert D^{-1}$ from the left. This yields a set of $K$
linear equations
\begin{equation}
\label{main_xk}
x_j  + \sum_{k=1}^K M_{jk} x_k = R_j \ ,
\end{equation}
\noindent
where
\begin{equation}
\label{M_R_def}
M_{jk} = \langle a_j \vert D^{-1} \vert b_k \rangle \ , \ \ R_j =
\langle a_j \vert D^{-1} \vert \Phi \rangle \ ,
\end{equation}
\noindent
and we have tacitly assumed that $D^{-1}$ exists. Once the unknown
quantities $x_k$ are found by solving the finite-dimensional set
(\ref{main_xk}) numerically, the vector $\vert \mu \rangle$ can be
easily computed according to
\begin{equation}
\label{mu_xk}
\vert \mu \rangle = D^{-1} \left( \vert \Phi \rangle - \sum_{k=1}^K
  x_k \vert b_k \rangle \right) \ .
\end{equation}
\noindent
It can be seen that the T-matrix is given by
\begin{equation}
\label{T_dir}
T = \sum_{j,k=1}^K \vert b_j \rangle M_{jk}^{-1} \langle a_k \vert \ .
\end{equation}
\noindent
Note that (\ref{mu_xk}) can be used to compute $\mu_n$ with arbitrary
$n$. In this sense, (\ref{mu_xk}) is truly the solution to the
infinite-dimensional system of equations.  The coefficients $M_{jk}$
and $R_j$ can be computed numerically or analytically, depending on
the problem. In what follows, we discuss computation of these
quantities for the particular case of star transform.

As was discussed in Sec.~\ref{subsec:trunc}, there exist two different
ways to truncate (\ref{main}), and here we use the particular
truncation in which $A$ is infinite (not truncated) while the
components $\Phi_n$ with $\vert n \vert > n_{\rm max}$ are set to
zero. Also, the matrix $D$ in the case of star transform is diagonal.
Correspondingly, the expression for $R_j$ (\ref{M_R_def}) contains
only finite sums, which can be computed without any approximations. We
also recall that the particular case of the star transform is obtained
if $\vert b_k \rangle = s_k \alpha_k \vert a_k \rangle$. Consequently,
the expression for $M_{jk}$ takes the form $M_{jk} = \langle a_j \vert
D^{-1} \vert a_k \rangle s_k \alpha_k$. This expression involves
infinite series in which all terms are known analytically. The series
can be easily summed numerically~\footnote{In the case $K=2$, the
  series can be summed analytically. We do not discuss this result
  because $M_{jk}$ can be computed numerically without any noticeable
  loss of precision for general $K$.}.  Convergence can be accelerated
by utilizing the known large-$n$ asymptote of the terms. Indeed, let
us write the series for the matrix element $ \langle a_j \vert D^{-1}
\vert a_k \rangle$ explicitly:
\begin{subequations}
\begin{align}
\label{M_kl_series}
 \langle a_j \vert D^{-1} \vert a_k \rangle &=
 \sum_{n=-\infty}^{\infty} \frac{1}{\beta_j + \kappa_n} 
\frac{1}{d_n} \frac{1}{\beta_k + \kappa_n} = \frac{1}{\beta_j \beta_k
  d_0} + \sum_{n=1}^\infty t_{jk,n} \ , \\
\label{t_kl_def}
t_{jk,n} &= \frac{1}{\beta_j + \kappa_n} \frac{1}{d_n} \frac{1}{\beta_k +
  \kappa_n} + \frac{1}{\beta_j - \kappa_n} \frac{1}{d_{-n}}
\frac{1}{\beta_k - \kappa_n} \ ,
\end{align}
\end{subequations}
\noindent
where $\beta_j$ are defined in (\ref{beta_xi_def}) and we have
accounted for $\kappa_{-n} = - \kappa_n$. We note that the factor
\begin{equation*}
\beta_j \beta_k d_0 = iq \frac{u_{jy} u_{ky}}{u_{jz} u_{kz}}
\sum_{l=1}^K \frac{s_l}{u_{ly}}
\end{equation*}
\noindent
turns to zero when $q=0$, but this case can be considered separately
as is described in Sec.~\ref{subsec:q=0} below. We can now consider
the large-$n$ asymptote of $t_{jk,n}$. To this end we use the
following expansion of $d_n$:
\begin{equation}
\label{d_n_exp}
d_n = i \left( \Sigma_1 \frac{1}{\kappa_n} + \Sigma_2
  \frac{q}{\kappa_n^2} + \ldots \right) \ , \ \ \vert n \vert
\rightarrow \infty \ ,
\end{equation}
\noindent
where 
\begin{equation}
\label{S_012_def}
\Sigma_0 = \sum_{k=1}^K
\frac{s_k}{\vert u_{kz} \vert} \ , \ \ \Sigma_1 = \sum_{k=1}^K
\frac{s_k}{u_{kz}} \ , \ \ \Sigma_2 = -\sum_{k=1}^K
\frac{s_k u_{ky}}{u_{kz}^2} \ .
\end{equation}
\noindent
The quantity $\Sigma_0$ is not used in this section but will be needed
below is Sec.~\ref{subsec:q=0}. From this, we obtain the following
expansion for $t_{jk,n}$:
\begin{equation}
\label{t_n_exp}
t_{jk,n} = \tau_{jk} \frac{1}{\kappa_n^2} + 
O \left(\frac{1}{\kappa_n^4} \right) \ , \ \  n \rightarrow \infty \ ,
\end{equation}
\noindent
where
\begin{equation}
\label{U_def}
\tau_{jk} = \frac{2i\left[\left( \beta_j + \beta_k \right)\Sigma_1 + q
    \Sigma_2 \right]}{\Sigma_1^2} 
= \frac{2iq}{\Sigma_1^2} \sum_{l=1}^K \frac{s_l}{u_{lz}}\left(
\frac{u_{jy}}{u_{jz}} + \frac{u_{ky}}{u_{kz}} - \frac{u_{ly}}{u_{lz}} \right) \ .
\end{equation}
\noindent
Note that the above derivations assume that $\Sigma_1 \neq 0$.  If
$\Sigma_1 = 0$, the diagonal elements of $D$ decay with $n$ as $1/n^2$
or faster, and the infinite system of equations (\ref{main}) with the
right-hand side coefficients $\Phi_n$ truncated to zero for $\vert n
\vert > n_{\rm max}$ does not have a solution. In
Sec.~\ref{subsec:q=0} below, we also show that the star transform with
$\Sigma_1=0$ is not invertible at $q=0$.

We can now use the result (\ref{U_def}) to rewrite (\ref{M_kl_series})
as
\begin{equation}
\label{M_kl_series_shift}
 \langle a_j \vert D^{-1} \vert a_k \rangle = \frac{1}{\beta_j \beta_k d_0} + \frac{L^2 \tau_{jk}}{24} +
\sum_{n=1}^\infty \left( t_{jk,n} -  \frac{\tau_{jk}}{\kappa_n^2} \right) \ .
\end{equation}
\noindent
The terms $(t_{jk,n} - \tau_{jk}/\kappa_n^2)$ decay with $n$ as
$1/n^4$ and, therefore, fast numerical convergence of
(\ref{M_kl_series_shift}) can be expected.

The disadvantage of the method presented here is that it is not
regularized and can become numerically unstable. However, $A^*A +
\lambda^2 {\mathbb I}$, similarly to $A$, contains one diagonal and a
finite number of separable terms. We can use this fact to generalize
the method for computation of the Tikhonov-regularized pseudo-inverse.

\subsection{Computational complexity}
\label{subsec:comp}

Each iteration of the methods of Sec.~\ref{subsec:A_pseudo} involves
either one or a few matrix-vector products.  Assuming for simplicity
$N \times N$ square $A$, we obtain the computational complexity of
$O(N^2)$ per iteration and the total computational complexity of
$O(KN^2)$. It should be emphasized that the matrix-vector products
involve a diagonal matrix at the first iteration but for the
subsequent iteration the diagonality is lost.  This is why we estimate
the computational complexity to be $O(N^2)$ per iteration rather than
$O(N)$.  We should also keep in mind that $A$ is parameterized by the
Fourier variable $q$ and that numerical inversion of $A(q)$ must be
performed for every value of $q$ used.  Consider an example of
reconstructing a rasterized $N\times N$-pixel image.  In this case,
the number of discrete values of $q$ to be used is also $N$. Then the
cumulative computational complexity of reconstructing the image by the
iterative methods is $O(KN^3)$. For comparison, a method based on
pixelization of the image and representing the star transform
(\ref{Star_Trans}) as a set of $N^2$ linear equations will result in
the computational complexity of $O(N^6)$ for direct methods or $O(M
N^4)$ for iterative methods such as the conjugate gradient descent,
where $M$ is the number of iterations needed for convergence.
Therefore, Fourier-space representation of the star transform and
iterative inversion of the resulting equations (parameterized by $q$)
can result in a very significant computational advantage.

An even larger computational advantage is gained by using the direct
method of Sec.~\ref{subsec:dir}. In this method the property of
diagonality in the matrix-vector products is not lost and the
computational complexity per one $q$ is $O({\rm max}(N,n_{\rm sum})) +
O(K^3)$, where $n_{\rm sum}$ is the maximum value of $n$ needed for
accurate approximation of the series of the type
(\ref{M_kl_series_shift}).  Assuming that $N \gg n_{\rm sum}, K^3$,
the cumulative computational complexity of reconstructing an image is
$O(N^2)$.

\section{Numerical stability}
\label{sec:stab}

In this section, we discuss the numerical stability of inverting the matrices
$A(q)$. Stability is understood here in the algebraic sense, namely,
we require the condition number of $A(q)$ (the ratio of its maximum
and minimum singular values) to be within the numerical range that
still allows for reliable numerical arithmetic. Simply stated, we say
that inversion of $A(q)$ is stable if the inverse of $A(q)$ can be
computed with numerical accuracy up to the usual round-off errors. 
This definition of stability is quite useful in practice, and several 
illustrative examples will be given below.
It should not be confused with the stability of inverting the star
transform in suitable function spaces, a topic which is deserving
of additional research.

Further, we will analyze the stability of inverting $A(q)$ for two
different cases: $qL \ll 1$ and $qL \gg 1$. Results for general $q$
are not available at this point, although we have seen numerical
evidence that $A(q)$ can be ill-posed at intermediate values of $q$
(that is, for $qL\sim 1$) even if it is well posed in the two cases
mentioned above. The instability at intermediate values of $q$ has
occurred in numerical experiments where all rays involved crossed the
same boundary. The corresponding artifacts were localized near the
boundary of the strip and were not very severe. We do not show such
examples in the Sec.~\ref{sec:sim} because the geometries with all
rays crossing the same boundary are not very important in practice.

The main concern for image reconstruction is instability at large
values of $q$. This instability results in high-frequency noise which
is not localized and tends to obscure the whole image. Examples of
such noise can be found, for example, in Ref.~\cite{florescu_11_1}.

We finally note that the Fourier reconstruction formula involves the
inverse of $A(q)$ inside an integral over $q$. Therefore, if inversion
of $A(q)$ is unstable in a very narrow interval of $q$ (that is,
$A^{-1}(q)$ has a small integral weight), such instability is not of
practical importance.

We now consider the two cases $qL\rightarrow 0$ and $qL\rightarrow
\infty$ separately.

\subsection{The case $qL \rightarrow 0$.}
\label{subsec:q=0}

The limit $q\rightarrow 0$ of (\ref{Phi_m_mu_n_1}) is not immediately
obvious; the cases $n=0$ and $n\neq 0$ must be considered separately
and, in the first case, the function $\alpha_k(q)$ must be expanded to
second order. The computations are, however, routine, and we adduce
here the final result. For $q=0$, (\ref{Phi_m_mu_n_1}) takes the
following form:
\begin{subequations}
\label{Phi_m_mu_n_q=0}
\begin{align}
\label{Phi_m_mu_n=0_q=0}
& \Phi_0 = \Sigma_0 \frac{\mu_0 L}{2} - i \Sigma_1 \sum_{m\neq 0}
\frac{\mu_m}{\kappa_m} \ , & n=0 \ , \\
\label{Phi_m_mu_n=/=0_q=0}
& \Phi_n = i\Sigma_1 \frac{\mu_n - \mu_0}{\kappa_n} \ , & n\neq 0 \ ,
\end{align}
\end{subequations}
\noindent
where $\Sigma_0$, $\Sigma_1$ and $\Sigma_2$ are defined in
(\ref{S_012_def}). 

In order for (\ref{Phi_m_mu_n_q=0}) to be invertible, the factors
$\Sigma_0$ and $\Sigma_1$ defined above must be nonzero
simultaneously. It is possible to select such ray geometries that this
condition holds. For example, consider the $K=4$ star with the
coefficient matrix $c_{jk}$ given in Sec.~\ref{sec:derivation}. In
this case $s_1=s_3=1$ and $s_2=s_4=-1$. If we take $u_{1z} = u_{3z} =
u/2$ and $u_{2z}=u_{4z} = - u$, where $0<u<1$, then $\Sigma_0 = 2/u$
and $\Sigma_1 = 6/u$.

If $\Sigma_0 \neq 0$ and $\Sigma_1 \neq 0$, we can invert
(\ref{Phi_m_mu_n_q=0}) analytically. Using the equality
$\sum_{m\neq 0}(1/\kappa_m) = 0$, we find that the solution to
(\ref{Phi_m_mu_n_q=0}) is
\begin{subequations}
\label{mu_n_q=0_sol}
\begin{align}
\label{mu_n=0_q=0_sol}
& \mu_0 = \frac{2}{L \Sigma_0} \sum_{m=-\infty}^\infty \Phi_m =
\frac{2 \tilde{\Delta}(0)}{\Sigma_0} \ , & n= 0 \ , \\
\label{mu_n=/=0_q=0_sol}
& \mu_n = \mu_0 - i \frac{\kappa_n \Phi_n}{\Sigma_1} \ , & n \neq 0 \ . 
\end{align}
\end{subequations}
\noindent
where $\tilde{\Delta}(q)$ is defined in (\ref{Delta_def}). We note
that the coefficients $\mu_n$ defined by (\ref{mu_n_q=0_sol}) can be
obtained by a term-by-term differentiation of a Fourier series of a
discontinuous function. The resultant series does not converge in the
usual sense but rather yields two delta-functions centered at $z=0$
and $z=L$. Therefore, the series should not be evaluated numerically
too close to the boundaries of ${\mathbb S}$. However, sufficiently
far from these boundaries, the series converges quite fast to the
correct result, that is, to the function $\tilde{\mu}(0,z)$, as can be
easily verified numerically using a number of examples. The region of
bad convergence is small and can be reduced in practice to one or two
pixels (on each side) of an image containing $\sim 100$ pixels in the
$Z$-direction. We emphasize that in order to obtain good convergence,
the coefficient $\mu_0$ must be determined with sufficient precision
from (\ref{mu_n=0_q=0_sol}).  Thus, (\ref{Phi_m_mu_n_q=0}) can be
solved in two steps. In the first step we compute $\mu_0$ according to
(\ref{mu_n=0_q=0_sol}).  This step is well-defined as long as
$\Sigma_0 \neq 0$.  In the second step we use the previous result to
compute $\mu_n$ for $n\neq 0$ according to (\ref{mu_n=/=0_q=0_sol}).
This step is well-defined if $\Sigma_1 \neq 0$.

\begin{figure}
\centerline{
\subfigure[]{\epsfig{file=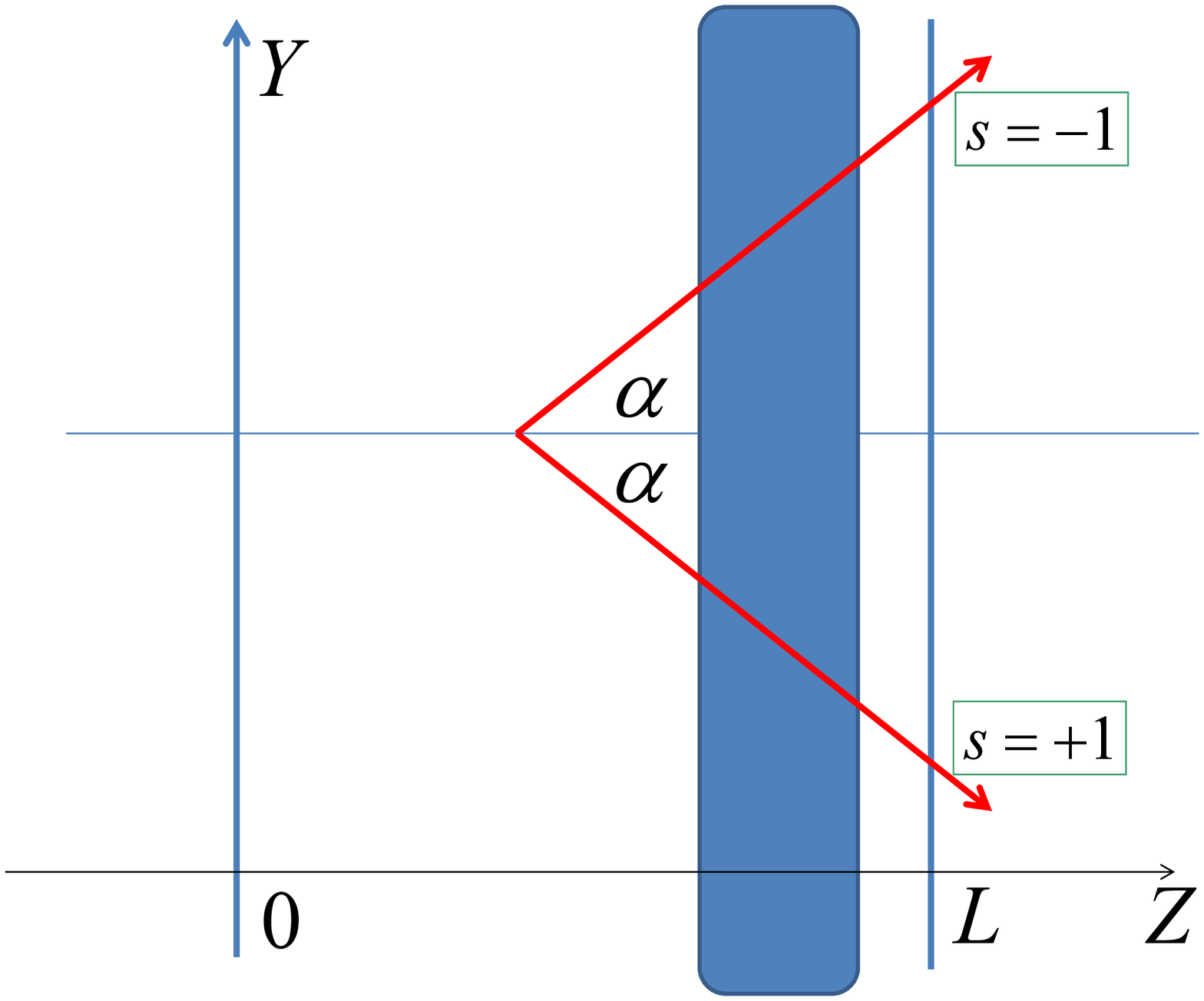,height=2.0in,clip=}}
\subfigure[]{\epsfig{file=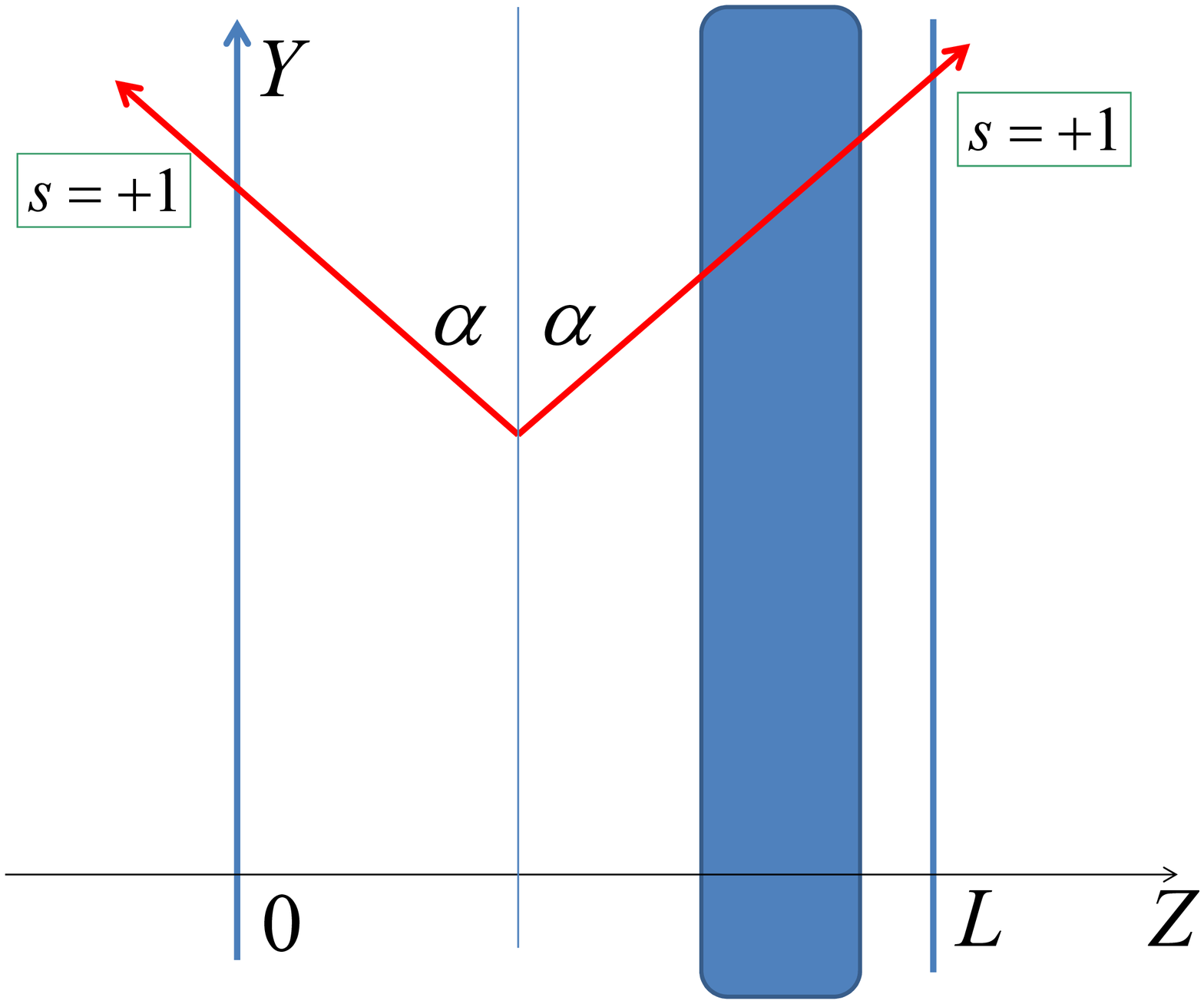,height=2.0in,clip=}}
}
\caption{\label{fig:symm} (color online) Two-ray symmetric stars with $\Sigma_1=0$. Here
  $\alpha$ denotes the angle. The long vertical inhomogeneity shown by
  the shaded rectangle is difficult to reconstruct using these
  geometries. We note that image reconstruction for the two cases
  shown in the figure is also ill-posed in the limit $qL\rightarrow \infty$ 
  because the number of rays is even 
 (see Sec.~\ref{subsec:q=Inf} below).}
\end{figure}

If $\mu_0$ is known or can be measured independently, then the
occurrence of $\Sigma_0 = 0$ does not pose a problem for image
reconstruction because only (\ref{mu_n=/=0_q=0_sol}) needs to be used
in this case. In Sec.~\ref{sec:proj}, we have explained how $\mu_0$
can be obtained using measurements of ballistic (non-scattered) rays.
However, the occurrence of $\Sigma_1=0$ is truly problematic. We will
refer to the imaging geometries with $\Sigma_1=0$ as to {\em
  symmetric}. We note that arrangements with $\Sigma_1=0$ have been
inadvertently used by us before, e.g., in \cite{florescu_10_1}. Image
reconstruction is still possible in this case, but it is not possible
to estimate correctly the integrals of the type $\int_{-\infty}^\infty
\mu(y,z) dy$~\footnote{Note that if $\Sigma_1=0$ but $\Sigma_0\neq 0$,
  it is still possible to estimate correctly $\mu_0(q)$ and, by
  Fourier transform, the quantity $\int_0^L \mu(y,z)dz$.}. This point
is illustrated graphically is Fig.~\ref{fig:symm} where we show two
star configurations with $K=2$ and $\Sigma_1=0$. The (a) configuration
has been used by us in~\cite{florescu_10_1,florescu_11_1}. It can be
seen that the long vertical inhomogeneity shown in the figures by a
dark-shaded rectangle is difficult to reconstruct using the ray
geometries shown in the figure. The data function in these two cases
is invariant with respect to the shift of the inhomogeneity along the
$Z$ axis as long as the rays do not intersect or touch the upper or
lower sides of the rectangle. The mathematical manifestation of this
observation is the ill-posedness of (\ref{Phi_m_mu_n_1}) at $q=0$.

\subsection{The case $qL \rightarrow \infty$}
\label{subsec:q=Inf}

Consider Eq.~(\ref{mu_xk}) in the limit $qL \rightarrow \infty$. It
can be seen that the second term inside the brackets is larger than
the first term by a factor $O(qL)$. Conversely, the second term in the
right-hand side of Eq.~(\ref{main}) is smaller than the first term by
a factor $O(1/qL)$. This is not surprising. The separable terms in the
expression for $A$ are due to the boundaries of ${\mathbb S}$. In the
high spatial frequency limit, the boundaries are unimportant and the
diagonal term in the expression for $A$ dominates the separable terms.
This fact can be used to analyze the stability of inverting $A(q)$
when $q$ is large. We emphasize that the high-frequency instability is
of primary concern for image reconstruction, as will be illustrated
with several numerical examples below.

Since in the limit considered the diagonal term $D$ dominates in
(\ref{mu_xk}), we will analyze the conditions under which the diagonal
elements $d_n(q)$, are nonzero for all values of $n$.
 
Let us write $d_n(q)$ in the form
\begin{equation}
\label{d_n_theta}
d_n(q) =  \sum_{k=1}^{K}\frac{is_k}{\hat{\bf{u}}_k \cdot(q,\kappa_n)} =
\frac{i}{\vert (q,\kappa_n) \vert} 
\sum_{k=1}^{K}\frac{s_k}{\hat{\bf u}_k \cdot \hat{\bf v}} \ , 
\end{equation}
\noindent
where $\hat{\bf v} = (q,\kappa_n) / \sqrt{q^2 + \kappa_n^2}$ is the
unit vector pointing in the direction of $(q,\kappa_n)$. Obviously,
$\hat{\bf v}$ is defined only if $(q,\kappa_n) \ne 0$, which is
obviously the case here. We then define the function
\begin{equation}
\label{f_def}
f(\theta) = \sum_{k=1}^{K} \frac{s_k}{\cos(\theta - \theta_k)} \ , 
\end{equation}
\noindent
where $\theta$ and $\theta_k$ are the polar angles of $\hat{\bf v}$
and $\hat{\bf u}_k$ in the $YZ$ plane. It is clear that $d_n(q)$ can
not turn to zero if $f(\theta)$ does not have zeros for real $\theta$.
On the other hand, if $f(\theta)$ has zeros, then we can find an
arbitrarily small element $d_n(q)$. Therefore, the sufficient and
necessary condition of stability of inverting the star transform is
that the function $f(\theta)$ (\ref{f_def}) does not have zeros on the
real axis. We note that $f(\theta + \pi) = - f(\theta)$ and,
therefore, it is sufficient to consider the interval $0 \leq \theta <
\pi$.

Of course, for any given set of $\hat{\bf u}_k$ and $s_k$, it is a
trivial matter to plot $f(\theta)$ and visually determine whether it
has zeros or not. We will, however, show that $f(\theta)$ always has
zeros if the number of rays $K$ is even and, moreover, if $K$ is odd,
then $f(\theta)$ has zeros if the vectors $s_k \hat{\bf u}_k$ can be
placed in the same half-plane (have simultaneously non-negative
projections onto the same axis).  Therefore, for inversion of the star
transform to be well-posed for $qL\gg 1$, the following three
necessary conditions must hold: (i) the number of rays $K$ should be
odd; (ii) the vectors $s_k \hat{\bf u}_k$ should not be contained the
same half-plane. This result will allow us to exclude star
configurations that are ill-posed {\em a priori}.

\begin{figure}
\centerline{
\subfigure[]{\rotatebox[]{-90}{\epsfig{file=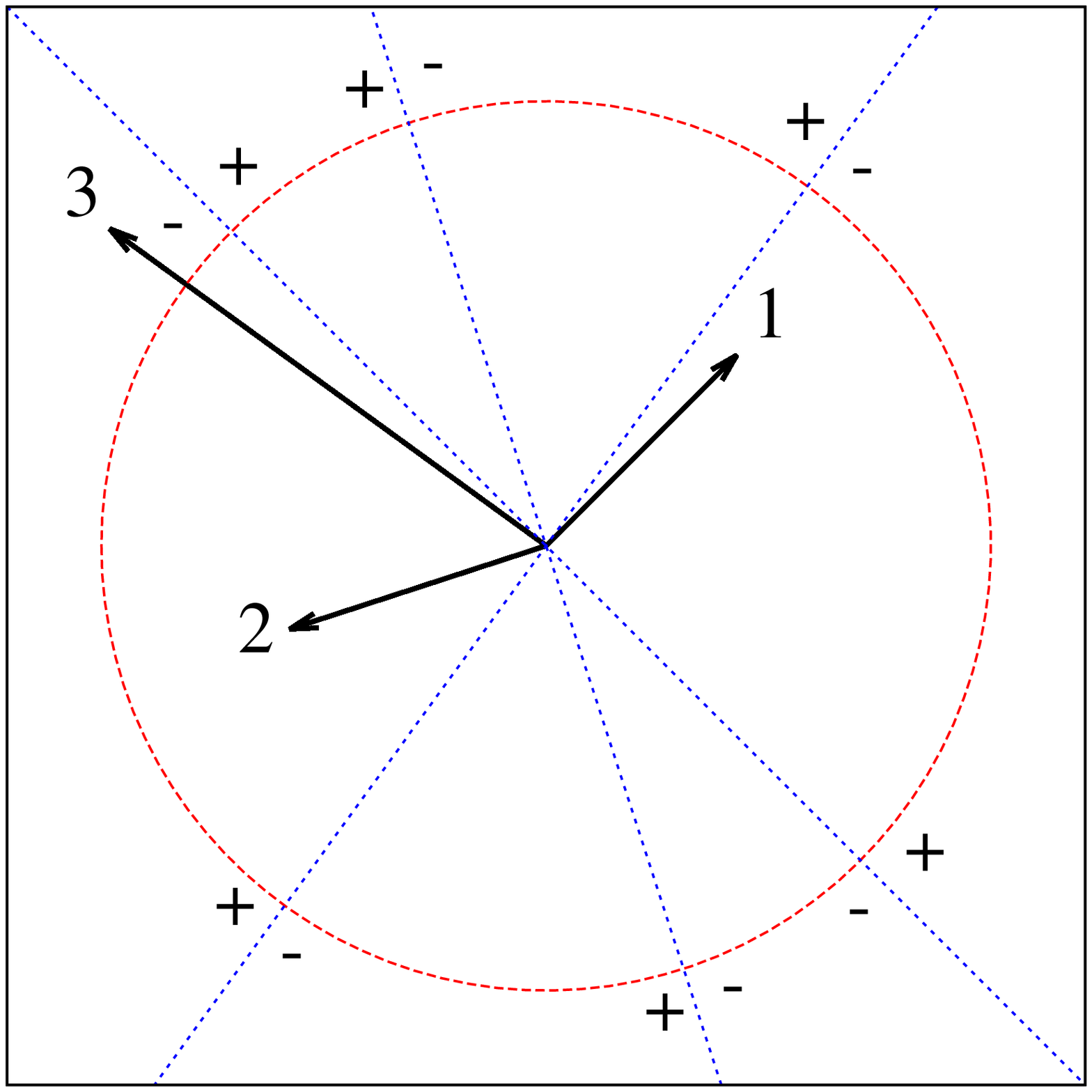,width=2.5in,bbllx=50bp,bblly=50bp,bburx=554bp,bbury=770bp,clip=}}}
\subfigure[]{\rotatebox[]{-90}{\epsfig{file=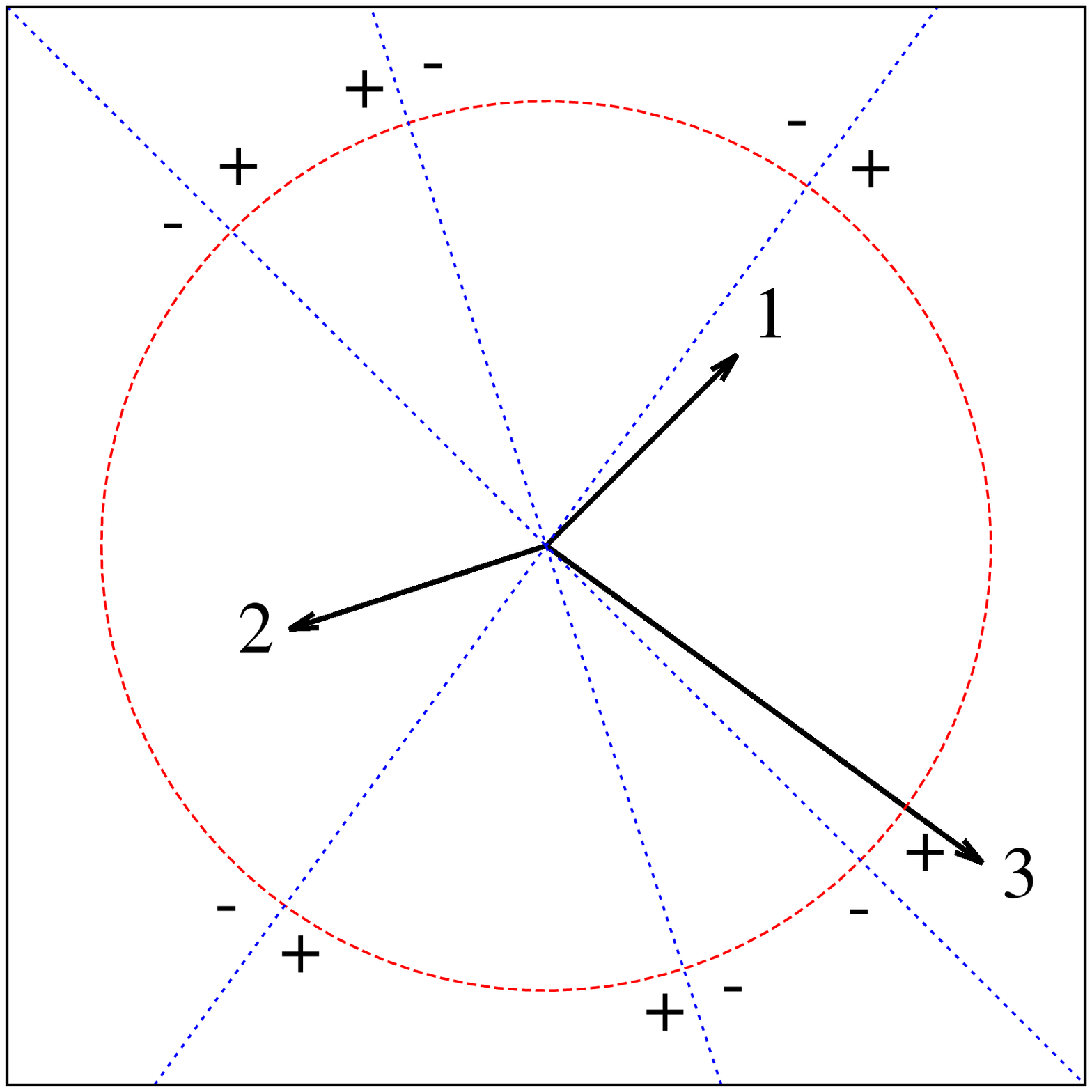,width=2.5in,bbllx=50bp,bblly=50bp,bburx=554bp,bbury=770bp,clip=}}}
}
\caption{\label{fig:dirs} (color online) Sign diagrams for the function $f(\theta)$ in the case 
  $K=3$ and $s_1=s_2=1$, $s_3=-2$. In both cases (a) and (b),
  $\theta_1 = 0.25\pi$ and $\theta_2 = 1.1\pi$, and the third angle is
  $\theta_3 = -0.2\pi$ (a) and $\theta_3 = 0.8$ (b). Angles are
  measured from the positive direction of the $Z$-axis in the
  counter-clockwise direction. It can be seen that the drawing of the
  third line, which runs from the top right to the bottom left corner,
  removes the contradiction in the case (b) but not in (a). Also, in
  (a) all vectors $s_k \hat{\bf u}_k$ can be placed in the same
  half-plane while in (b) the same is not true.}
\end{figure}

We will now show how the above conditions have been obtained. We start
with the observations that, when $\theta = \theta_k \pm \pi/2$ (that
is, when $\hat{\bf v}$ is perpendicular to one of the unit vectors
$\hat{\bf u}_k$), $f(\theta)$ diverges, and that $f(\theta_k -
\epsilon)$ and $f(\theta_k + \epsilon)$ have different signs, where
$\epsilon$ is an infinitesimal constant. Between the singular points,
$f(\theta)$ is continuous. We therefore must determine whether
$f(\theta)$ changes sign in at least one of the intervals where it is
continuous. To this end, diagrams such as those shown in
Fig.~\ref{fig:dirs} can be useful.  Consider a circle and draw each
vector $s_k \hat{\bf u}_k$ as an arrow originating from the circle
center. Then draw a perpendicular to each arrow, also through the
circle center, as is shown by the dashed lines. Near each point were a
dashed line intersects the circumference, draw a pair of signs, plus
and minus, on each side of the line. These signs indicate the sign of
$f(\theta)$ near the singularity. There are two singularities and four
signs associated with each line.  When placing the signs, a couple of
obvious rules must be obeyed: different signs are placed on different
sides of a singularity and similar signs are placed in each of the two
half-planes created by a given line. Note that changing the sign of
$s_k$ corresponds to changing all signs associated with the
corresponding line. Now, let us, starting from an arbitrary point,
move along the circumference in any direction and make a complete
revolution. Crossing the dashed lines corresponds to crossing the
singularities of $f(\theta)$ while motion from one dashed line to the
next corresponds to the intervals of $\theta$ where $f(\theta)$ is
continuous. If the motion over an interval of continuity connects two
opposite signs, as is the case in the top and bottom segments of the
diagram in Fig.~\ref{fig:dirs}(a), then $f(\theta)$ has at least one
zero in that interval.

First, let us consider the case of even $K$. Assume that $f(\theta)$
does not have zeros. Then take an arbitrary line and place the signs
associated with it in accordance with the above two rules but
otherwise arbitrarily, and start moving from the two singular points
just ``signed'' in, say, the clock-wise direction. The hypothesis that
$f(\theta)$ does not have zeros forces a unique choice of signs for
the next two singular points. Continue this process until only one
``unsigned'' line remains. If $K$ is even, then this last line can not
be ``singed'' without violating the underlying hypothesis. There will
appear two symmetrically-situated intervals of $\theta$ (contained
between two lines) where $f(\theta)$ is continuous and changes sign.
Therefore, we have arrived at a contradiction and $f(\theta)$ must
have zeros. For example, if $K=2$, $f(\theta)$ has zeros when $\theta
= \pm \Theta[s_2 \hat{\bf u}_1 + s_1 \hat{\bf u}_2]$, where the last
expression denotes the polar angle of the vector in the square
brackets.

Consideration of odd $K$ is somewhat more complicated. We can again
make the hypothesis that $f(\theta)$ does not have zeros and start
from $K-1$ rays, where $K-1$ is even. As was discussed above, there
will appear two intervals of $\theta$ where $f(\theta)$ is continuous
and changes sign. Since these intervals are contained between the same
two lines, each of them can be divided in two by drawing one
additional, $K$-th line. However, the contradiction is removed only
for one particular choice of signs associated with this last line.
This is illustrated in Fig.~\ref{fig:dirs} for $K=3$. Here we start
with $K-1=2$ and break the two intervals of $\theta$ that connect
opposite signs with a third line. In the considered example, the
choice of signs for this third line depends on the direction of
$\hat{\bf u}_3$ [the coefficient $s_3$ is the same in cases (a) and
(b)]. In the case (b), the contradiction to the original hypothesis is
removed but in the case (a) it is not. It can be further determined by
inspection that in the case (a) all three vectors $s_k \hat{\bf u}_k$
(shown by arrows) are contained in the same half-plane.  More
generally, we can follow similar considerations for an arbitrary $K$
to show that the $K$-th ray removes the contradiction only if $s_K
\hat{\bf u}_K$ is not contained in the same half-plane as the vectors
$s_1 \hat{\bf u}_1, \ldots, s_{K-1} \hat{\bf u}_{K-1}$.

\begin{figure}
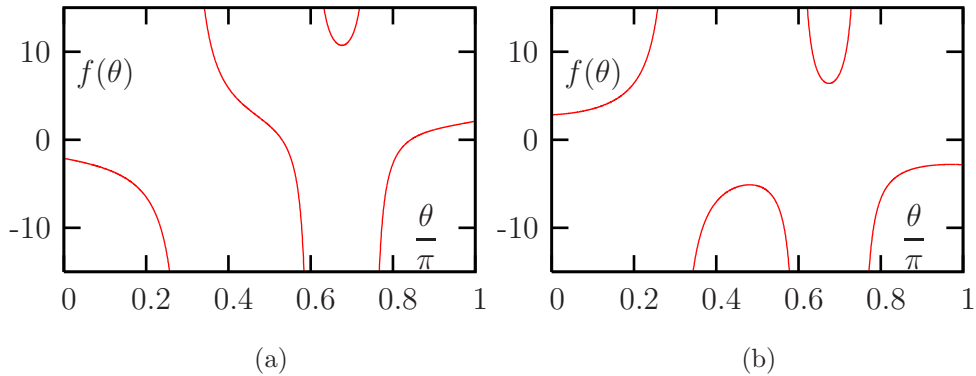

\centerline{
\subfigure[]{\input{Fig_4a.tex}}
\subfigure[]{\input{Fig_4b.tex}}
}
\caption{\label{fig:ftheta} (color online) 
  The function $f(\theta)$ for the two cases (a) and (b) shown in
  Fig.~\ref{fig:dirs}.}
\end{figure}

The function $f(\theta)$ for the two cases shown in
Fig.~\ref{fig:dirs} is plotted in Fig.~\ref{fig:ftheta} in the
interval $0\leq \theta \leq \pi$. Image reconstruction for the two
star configurations of Fig.~\ref{fig:dirs} are shown below in
Fig.~\ref{fig:Case_3}.

\begin{figure}
\centerline{\input{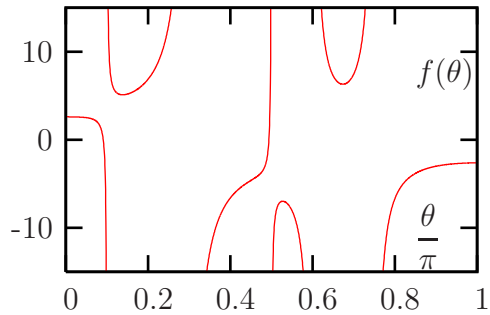}}
\caption{\label{fig:counter_example} (color online) A counter-example to conditions
  (i)-(iii) of Sec.~\ref{subsec:q=Inf}. Function $f(\theta)$ for the same
  rays as in Fig.~\ref{fig:dirs}(b) {\em plus} two additional rays
  with $s_4=-0.1$, $s_5=0.1$ and $\theta_4=\pi$, $\theta_5=0.6\pi$.
  The total number of rays is $K=5$. Conditions (i)-(iii) of
  Sec.~\ref{subsec:q=Inf} are satisfied but $f(\theta)$ has zeros.}
\end{figure}

We emphasize that the condition based on zeros of $f(\theta)$ is
necessary and sufficient. However the conditions (i)-(ii) of this
section are necessary only. We can easily find a case when (i)-(ii)
are satisfied but $f(\theta)$ has zeros. A counter-example of this
type for $K=5$ is shown in Fig.~(\ref{fig:counter_example}). Here we
have used the same three rays as in Fig.~\ref{fig:dirs}(b) plus two
additional rays with $s_4=-0.1$, $s_5=0.1$ and $\theta_4=\pi$,
$\theta_5=0.6\pi$. In this case the conditions (i)-(ii) are obviously
satisfied and, additionally, we have preserved the condition $\sum_k
s_k =0$, which is required for simultaneous reconstruction of
attenuation and scattering. 

Counter-examples of this type can be generated by considering a star
configuration with $K-2$ rays, where $K-2$ is odd, $\sum_{k=1}^{K-2}
s_k = 0$, and in which $f(\theta)$ has no zeros. We then add to the
configuration a pair of rays with the coefficients $s_{K-1} = - s_K$
and $\vert s_K \vert \ll 1$ to obtain a new configuration with $K$
rays. Addition of such ray pairs with a ``wrong'' sign of $s_K$ can
generate zeros in $f(\theta)$ without violating any of the conditions
(i)-(ii) or the sum rule $\sum_k s_k=0$. However, such
counter-examples are of little practical concern. Indeed, in all such
examples, the derivative of $f(\theta)$ near its zeros is very large.
This can be clearly seen in Fig.~\ref{fig:counter_example}. As was
already mentioned, the inverse of $A(q)$ appears in an integral over
$q$. Therefore, singularities of this sort have small integral weight
and do not result in noticeable image distortions or artifacts. Of
course in the limit $s_K \rightarrow 0$, the two additional rays do
not influence image reconstruction at all even though $f(\theta)$
still formally has zeros.

\section{Simulations}
\label{sec:sim}

In this section, we report several numerical experiments, which
illustrate possible applications of the star transform to imaging.  We
use different star geometries some of which are applicable to
simultaneous reconstruction of scattering and attenuation
coefficients. However, at this stage, we only show reconstructions of
the attenuation. Physically, the reconstructed quantity is $\delta
\mu(y, z)$, the deviation of the total attenuation coefficient from
its background value [see the discussion around
Eqs.~(\ref{Star_Trans_delta})]. As such, the reconstructed quantity is
zero outside of a few finite objects. Strictly speaking, the data
function used is $\delta\Phi$ defined in (\ref{Star_Trans_delta}).
However, as elsewhere in the paper, we refer to the reconstructed
quantity as to $\mu$ and to the data function as to $\Phi$.

We use Eq.~(\ref{Star_Trans}) to generate the data $\Phi({\bf R})$ by
computing the ray integrals analytically, which can be regarded as
``inverse crime'' (that is, generating the data from the same model as
is used for image reconstruction). We make our numerical tests more
realistic by including Poissonian random noise in the data. The noise
is added as follows. First, we recall that the data function $\Phi$ is
constructed as a linear superposition of functions $\phi_{jk}$, which,
in turn, are related the physical measurements of intensity $W_{jk}$
by (\ref{phi_jk_def}). We therefore add noise directly to the
functions $\phi_{jk}$ and then construct the data function of the star
transform, $\Phi$, according to (\ref{sum_3}). In this expression, the
numerical coefficients $c_{jk}$ are known precisely and determined as
described in Sec.~\ref{sec:derivation} but the terms $\phi_{jk}$
contain noise. To add noise to $\phi_{jk}$, we use the following
procedure. For each data point $\phi_{jk}$, we first compute an
integer $\bar{\mathcal M}$ according to $\bar{\mathcal M} = {\rm
  nint}[{\mathcal N} \exp(-\phi_{jk})]$, where ${\mathcal N}$ is a
fixed integer number and ${\rm nint}(x)$ denotes the nearest integer
to $x$. We then choose randomly a number ${\mathcal M}$ from a
Poissonian distribution whose average is $\bar{\mathcal M}$. Finally,
we compute the noise-affected data point $\phi_{jk}^\prime$ as
$\phi_{jk}^\prime = - \log( {\mathcal M}/{\mathcal N})$. It can be
seen that $\phi_{jk}^\prime \rightarrow \phi_{jk}$ in the limit
${\mathcal N} \rightarrow \infty$. This method of adding noise is
somewhat {\em ad hoc}, but it has a well controlled behavior and
distorts randomly the reconstructed images when ${\mathcal N}$ is not
very large. Undoubtedly, more realistic noise models will be required
in the future.

In all reconstructions, the data function and the reconstructed image
were sampled on a square grid with the step $h$, where $h=L/126$ ($L$
being the transverse width of the strip). Thus, there were $N=125$
real-space samples of $\Phi(Y,Z)$ in the $Z$-direction, and similarly
in $Y$. This corresponds to the truncation of Fourier coefficients
$\mu_n(q)$ such that $-n_{\max} \leq n \leq n_{\rm max}$, where
$n_{\rm max} = (N-1)/2 = 62$. The Fourier variable $q$ was similarly
sampled and truncated. Correspondingly, we considered a square matrix
$A$ of the size $N \times N$.

\begin{figure}
\centerline{
\epsfig{file=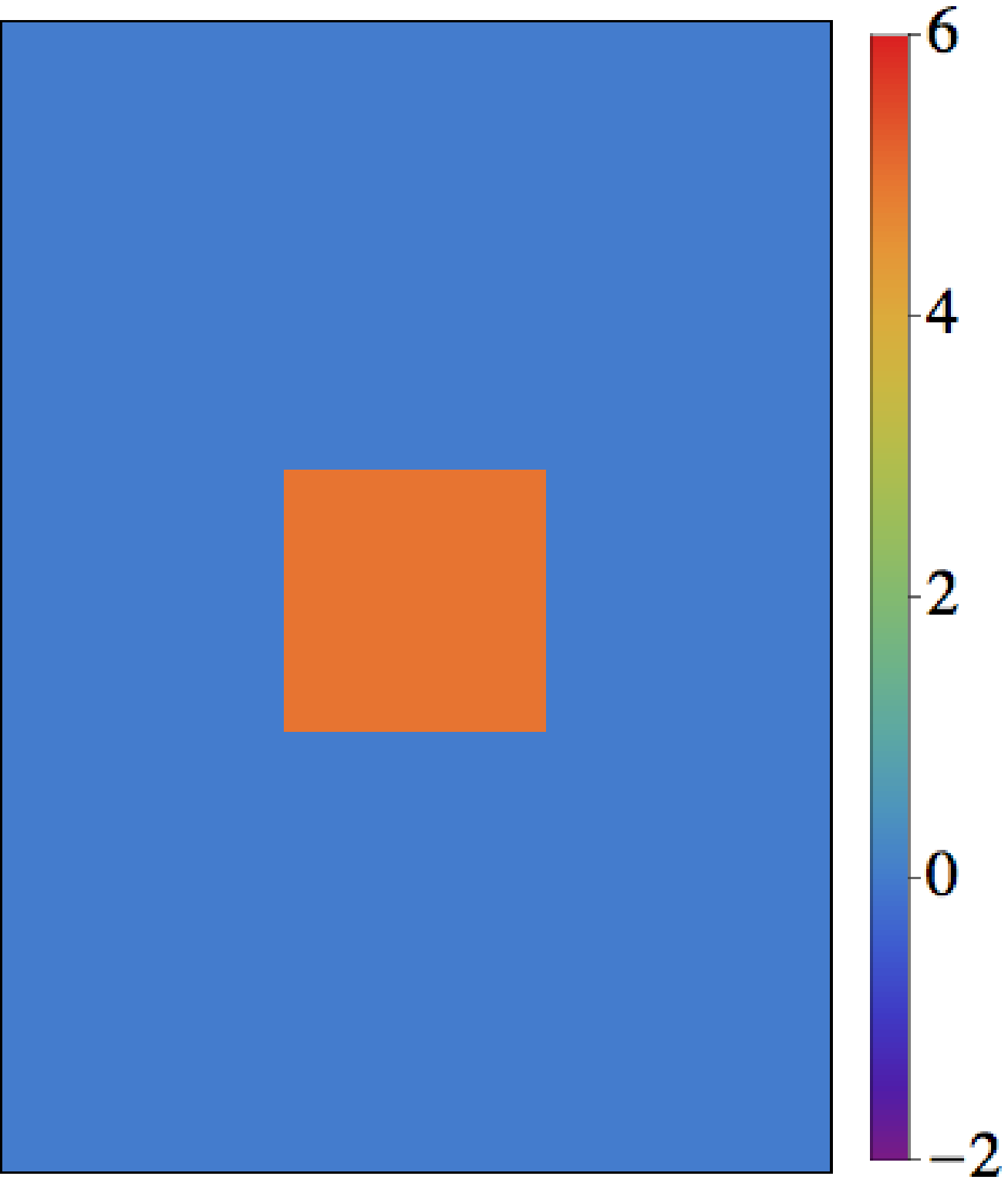,width=1.3in,bbllx=0bp,bblly=0bp,bburx=450bp,bbury=480bp,clip=}
\epsfig{file=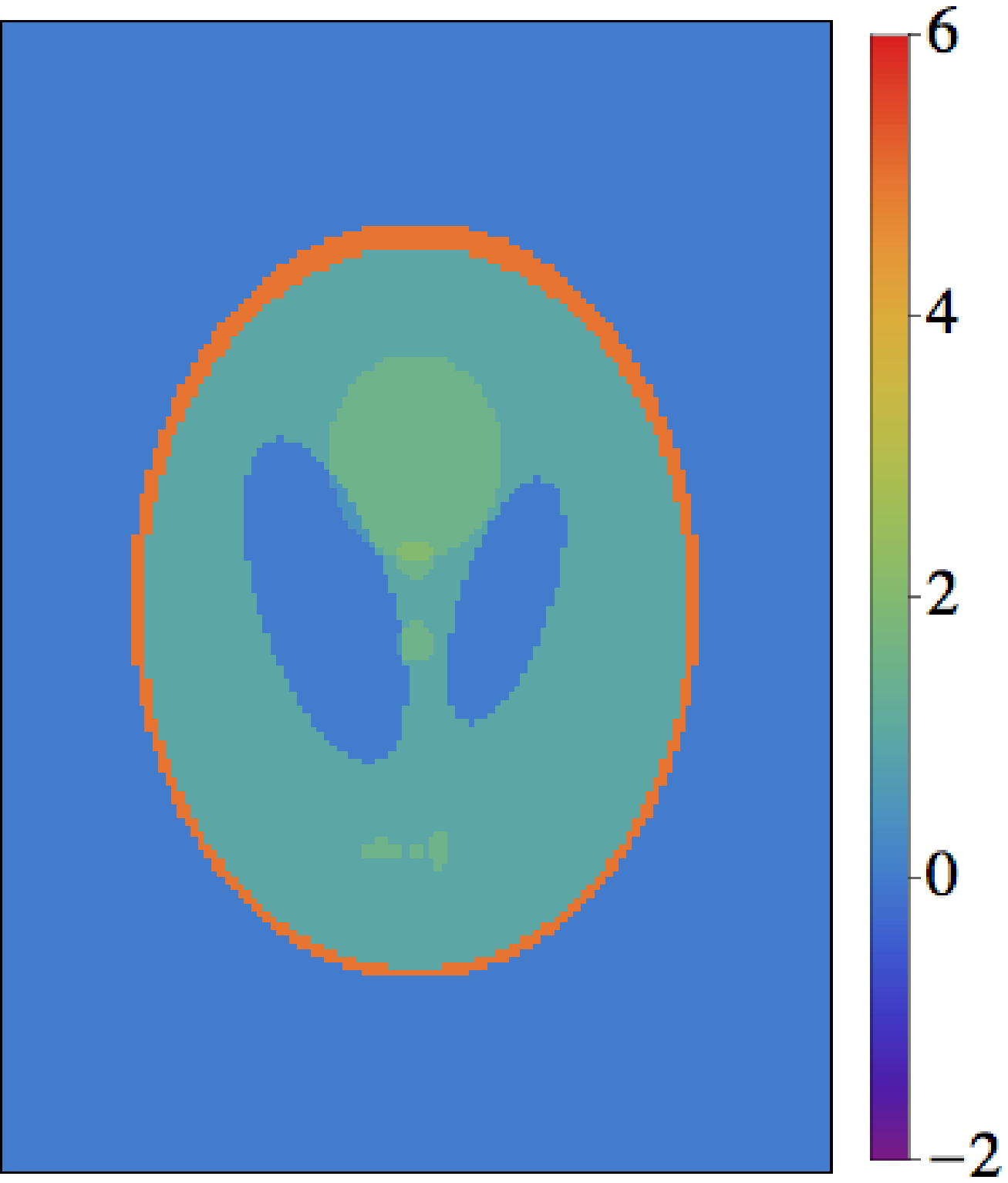,width=1.3in,bbllx=0bp,bblly=0bp,bburx=450bp,bbury=480bp,clip=}
}
\caption{\label{fig:phantoms} (color online) The phantoms used in
  the numerical experiments: square (left) and Shepp-Logan phantom
  (right). The quantity depicted by the color scale is the
  dimensionless parameter $\mu L$. Outside of the inhomogeneities,
  this quantity is set to zero. The same color scale as in this Figure
  is used in all reconstructed images shown below. Whenever the
  reconstructed values are outside of the interval $-2 < \mu L < 6$,
  we use the following color clipping: the areas with $\mu L<-2$ are
  shown by black color and the areas with $\mu L > 6$ are shown by
  white (this clipping is not used in this rendering of phantoms but
  will appear in the reconstructed images below).}
\end{figure}

To generate forward data, we employ two different phantoms: a square
and the Shepp-Logan phantom. The attenuation coefficient of the square
phantom was $\mu = 5/L$ and in the case of Shepp-Logan phantom it
varied from $\mu = 1/L$ to $\mu = 5/L$ (inside the inhomogeneities).
In the background, we had $\mu=0$. The phantoms are shown in
Fig.~\ref{fig:phantoms}.

Inversion of the star transform with two and three rays has been
simulated. The exact parameters of the imaging geometries used are
listed in Table~\ref{tab:1}, where we also list the number of zeros of
the function $f(\theta)$ in the interval $0\leq \theta \leq \pi$ and
the quantities $\Sigma_0$, $\Sigma_1$ (\ref{S_012_def}) for each
geometry used. As was discussed in the paper, these quantities can be
used to ascertain the ill-posedness of the star transform. The imaging
geometries are grouped into three cases. In Cases 1 and 2, all
coefficients $s_k$ are equal to unity.  As was discussed in
Sec.~\ref{sec:derivation}, these types of star transform are
applicable to reconstruction of purely absorptive contrast in a
uniform scattering background. The geometry of Case 3b satisfies all
the conditions (i)-(iv) of Sec.~\ref{sec:derivation} and therefore can
be used to reconstruct simultaneously attenuation and scattering
contrast. We note that the star transform of Case 3b can be obtained
from physical measurements as is suggested by the first coefficient
table in the end of Sec.~\ref{sec:derivation}.

\begin{table}
\centerline{
\begin{tabular}{|r|r|r|r|r|r|r|}
\hline
Case           &  1a   & 1b                & 2a      & 2b               & 3a      & 3b               \\
\hline
K              & \multicolumn{2}{|c|}{2}   & \multicolumn{2}{|c|}{3}    & \multicolumn{2}{|c|}{3}    \\
\hline
$s_1$          & \multicolumn{2}{|c|}{1}   & \multicolumn{2}{|c|}{1}    & \multicolumn{2}{|c|}{1}    \\
\hline
$s_2$          & \multicolumn{2}{|c|}{1}   & \multicolumn{2}{|c|}{1}    & \multicolumn{2}{|c|}{1}    \\
\hline
$s_3$          & \multicolumn{2}{|c|}{N/A} & \multicolumn{2}{|c|}{1}    & \multicolumn{2}{|c|}{-2}   \\
\hline
$\theta_1/\pi$ & 1       & 0.82            & \multicolumn{2}{|c|}{1}    & \multicolumn{2}{|c|}{0.25} \\
\hline
$\theta_2/\pi$ & 0.25    & 0.23            & \multicolumn{2}{|c|}{0.25} & \multicolumn{2}{|c|}{1.1}  \\
\hline
$\theta_3/\pi$ & \multicolumn{2}{|c|}{N/A} & -0.25   & -1/6             & -0.2    & 0.8              \\
\hline
NZ             & 1       & 1               & 0       & 0                & 2       & 0                \\  
$\Sigma_0$     & 2.41    & 2.52            & 3.83    & 3.57             & -0.01   & -0.01            \\ 
$\Sigma_1$     & 0.41    & 0.15            & 1.83    & 1.57             & -2.11   & 2.83           \\
\hline
\end{tabular}
}
\caption{\label{tab:1} Weight coefficients $s_k$ and ray angles 
$\theta_k$ and  for all cases considered in Sec.~\ref{sec:sim}. 
The angles are measured with respect to the positive direction of the $Z$-axis
(crossing the strip from left to right and shown by the dashed line in
Fig.~\ref{fig:phantoms}). Counter-clockwise rotation direction
is assumed to be positive. Also shown for each case are the number of
zeros NZ of the function $f(\theta)$ (\ref{f_def}) in the interval 
$0\leq \theta \leq \pi$ and the expansion coefficients
$\Sigma_0$ and $\Sigma_1$ (\ref{S_012_def}) rounded off to three significant figures. 
The Cases 3a and 3b correspond to the cases (a) and (b) of
Figs.~\ref{fig:dirs} and \ref{fig:ftheta}.} 
\end{table}

\subsection{Case 1}

\begin{figure}
\centerline{
\subfigure[No noise, $\lambda = 10^{-2}$]{
\epsfig{file=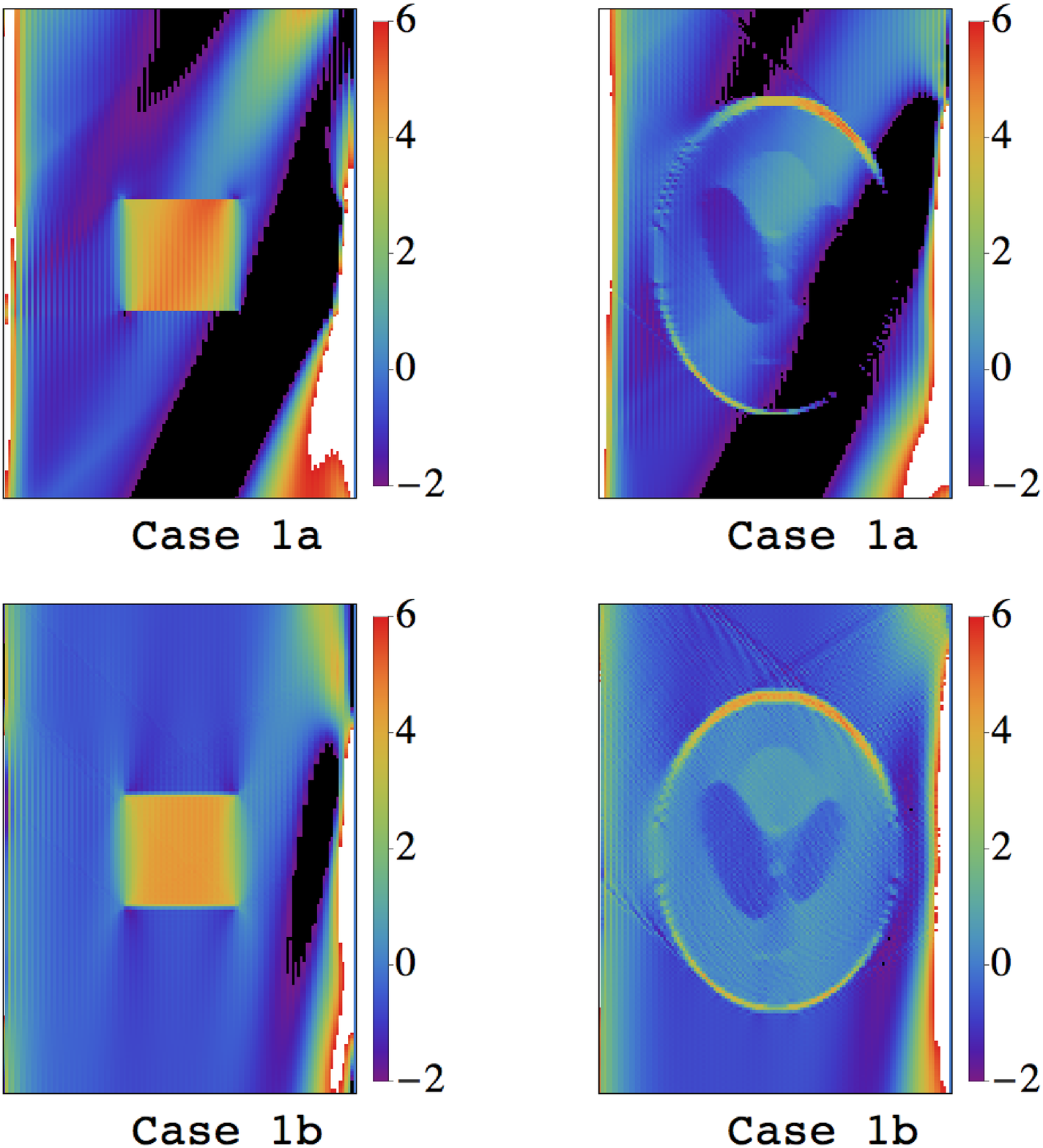,width=2.6in}}
\subfigure[${\mathcal N} = 4 \times 10^4$, $\lambda=10^{-2}$]{
\epsfig{file=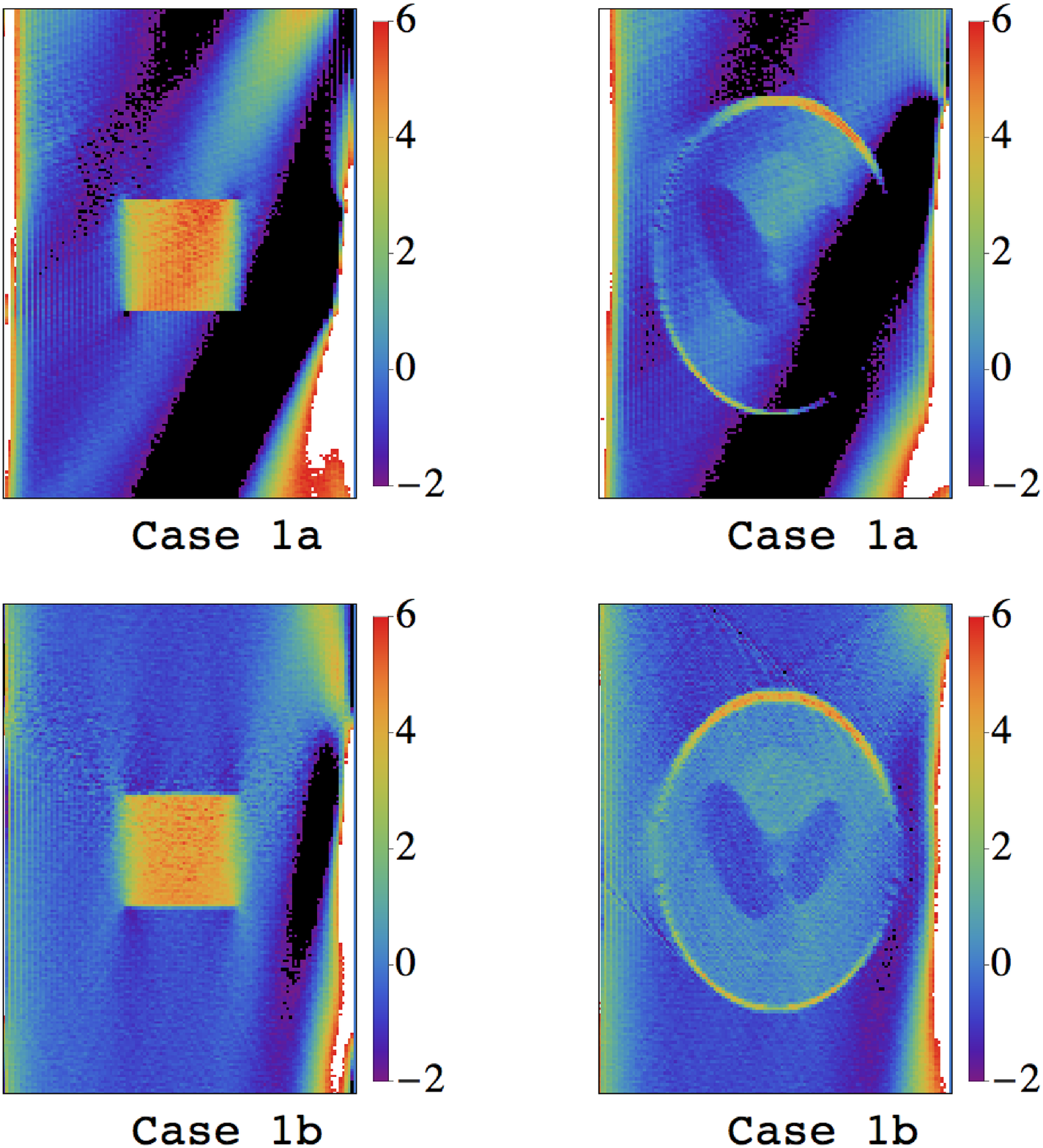,width=2.6in}}
}
\centerline{
\subfigure[${\mathcal N} = 10^4$, $\lambda=10^{-2}$]{
\epsfig{file=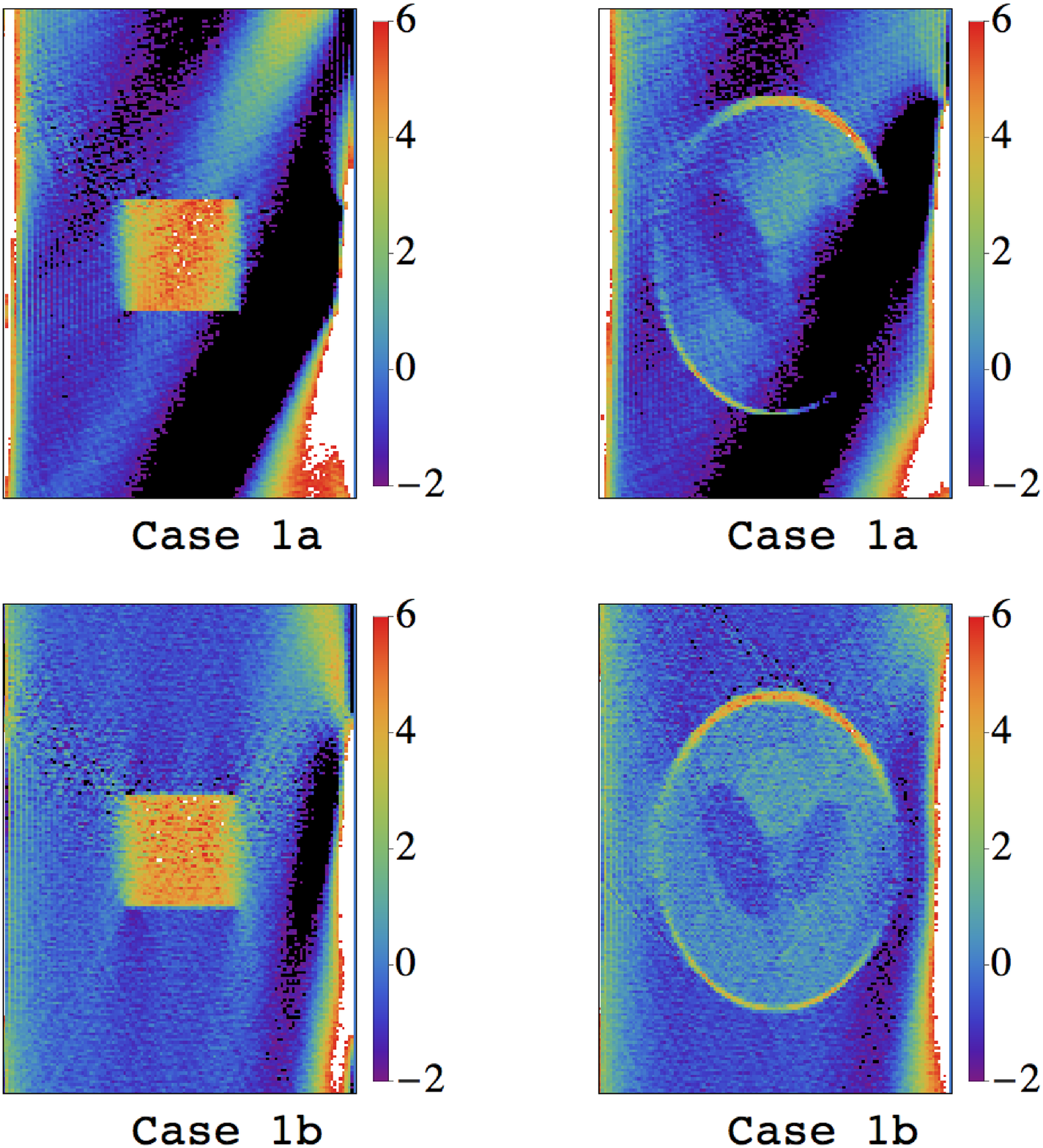,width=2.6in}}
\subfigure[${\mathcal N} = 2.5\times 10^3 $, $\lambda=10^{-2}$]{
\epsfig{file=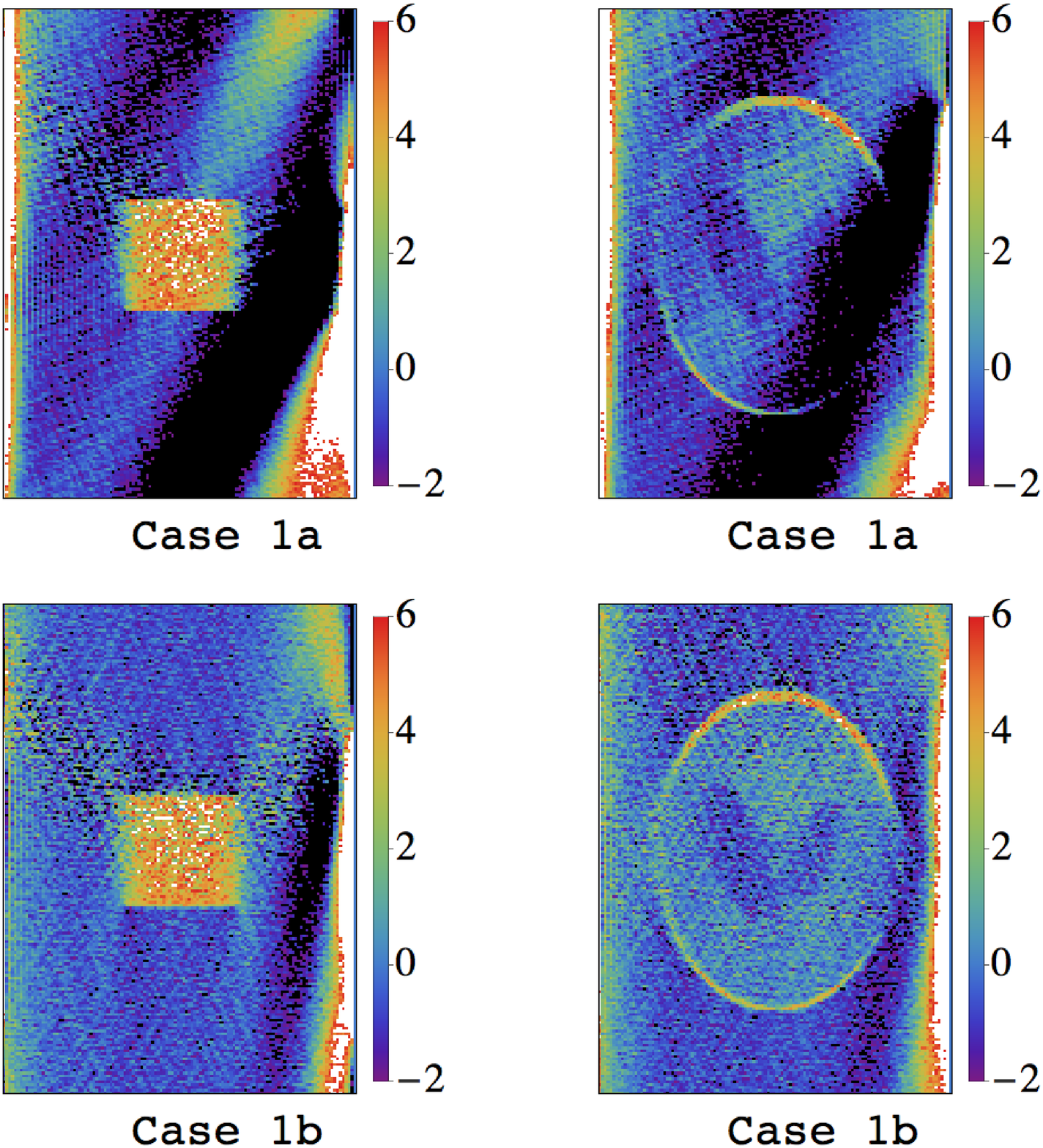,width=2.6in}}
}
\caption{\label{fig:Case_1} (color online) Case 1: Image reconstructions
  for various levels of noise and regularization parameter $\lambda$,
  as labeled. Here ${\mathcal N}$ is the integer parameter controlling the
  level of Poissonian noise in the data. the following clipping of the
  color scale has been used: Black color corresponds to numerical
  values $\mu L < -2$ and white color corresponds to $\mu L > 6$.}
\end{figure}

We start with image reconstructions for the ray geometries of Case
1 ($K=2$; see Tab.~\ref{tab:1} for more detail). Case 1 is only
applicable to reconstructing absorptive contrast in a medium with a
spatially-uniform scattering coefficient. Results for four different
levels of noise are shown in Fig.~\ref{fig:Case_1}. The images were
obtained by Tikhonov-regularized pseudo-inverse as described in
Sec.~\ref{subsec:A_pseudo}.

\begin{figure}
\centerline{\epsfig{file=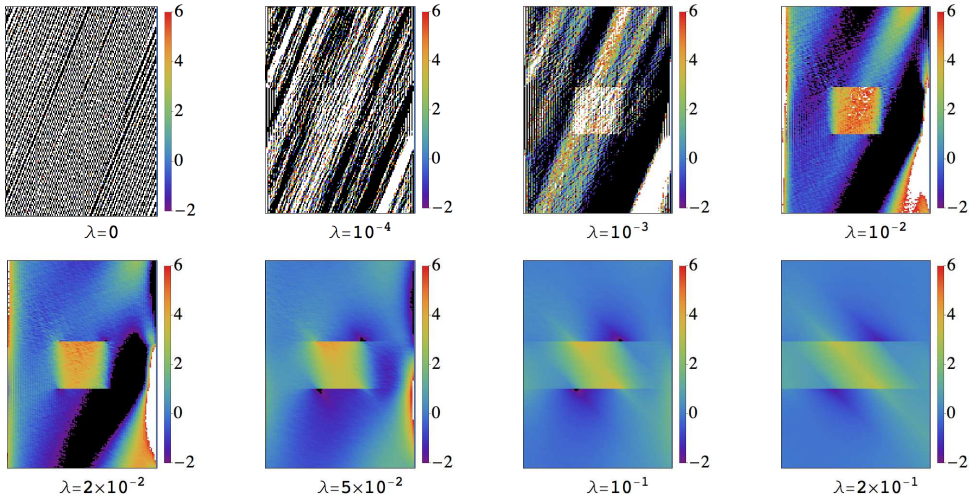,width=5.3in}}
\caption{\label{fig:Case_1_reg} (color online) 
  Case 1a: Effects of regularization for the noise level ${\mathcal N}=10^4$
  and different regularization parameters $\lambda$, as labeled.}
\end{figure}

The reconstructions contain fairly severe artifacts, especially in
Case 1a. This is a consequence of the ill-posedness of the star
transform at large values of $q$, in agreement with Sec.~\ref{subsec:q=Inf} 
($K$ is even in Case 1). Similar artifacts have been observed by us previously in the
purely numerical reconstructions utilizing two-ray
geometries~\cite{florescu_09_1,florescu_10_1}. The artifacts in Case
1b are less severe. We attribute this to the fact that the derivative
of $f(\theta)$ near the point where it turns to zero is much larger in
Case 1b than in Case 1a. Correspondingly, the singularity in Case 1a
has a larger integral weight. The role of the derivative of
$f(\theta)$ near its zeros was briefly discussed in the end of
Sec.~\ref{subsec:q=Inf}. Note that in Case 1, regularization was required
to obtain a recognizable image at all noise levels considered,
including the case with no noise. An illustration of the effects of
regularization is given in Fig.~\ref{fig:Case_1_reg}.

\subsection{Case 2}

\begin{figure}
\centerline{
\subfigure[No noise, $\lambda= 0$]{
\epsfig{file=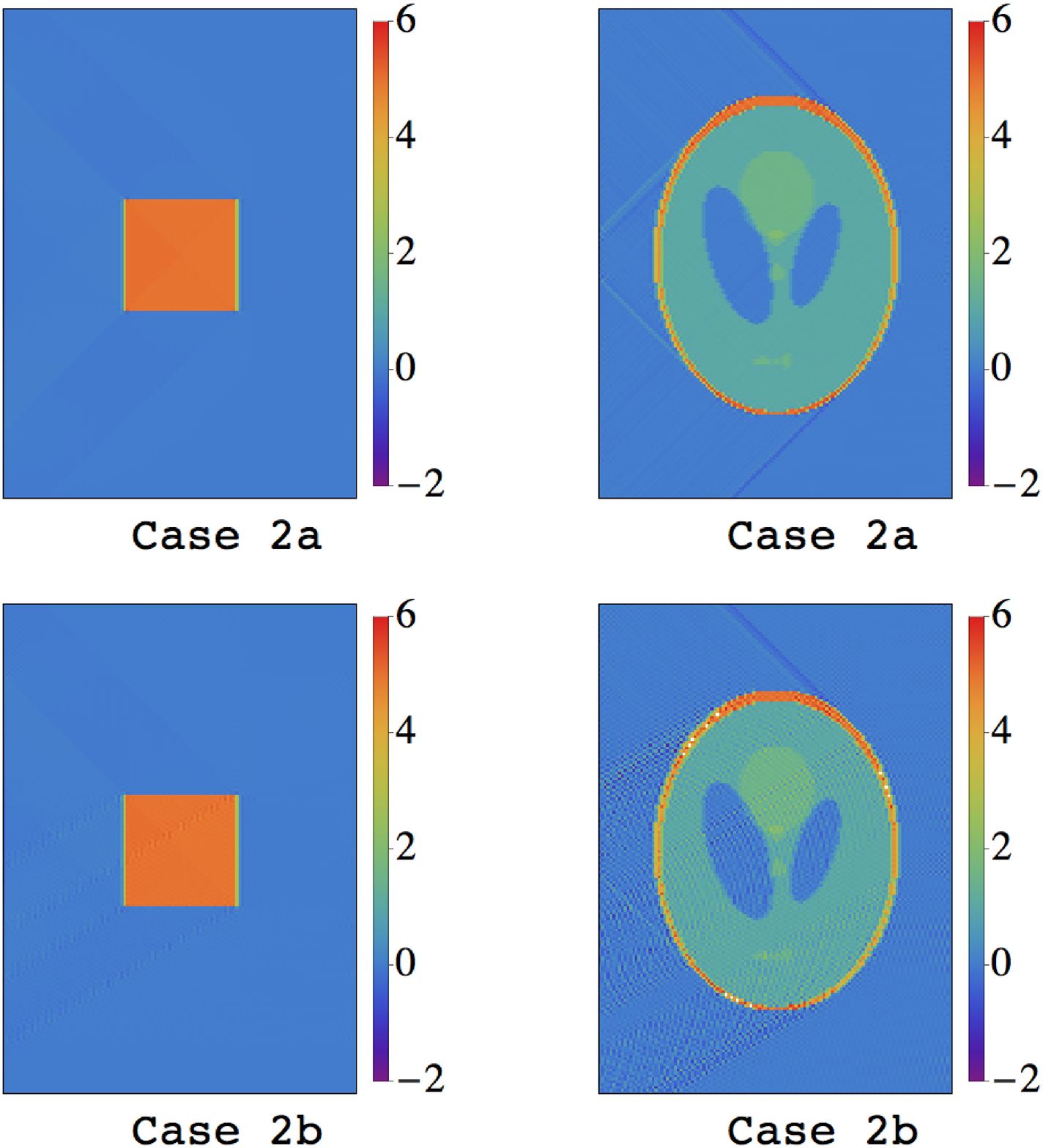,width=2.6in}}
\subfigure[${\mathcal N} = 4 \times 10^4$, $\lambda= 0$]{
\epsfig{file=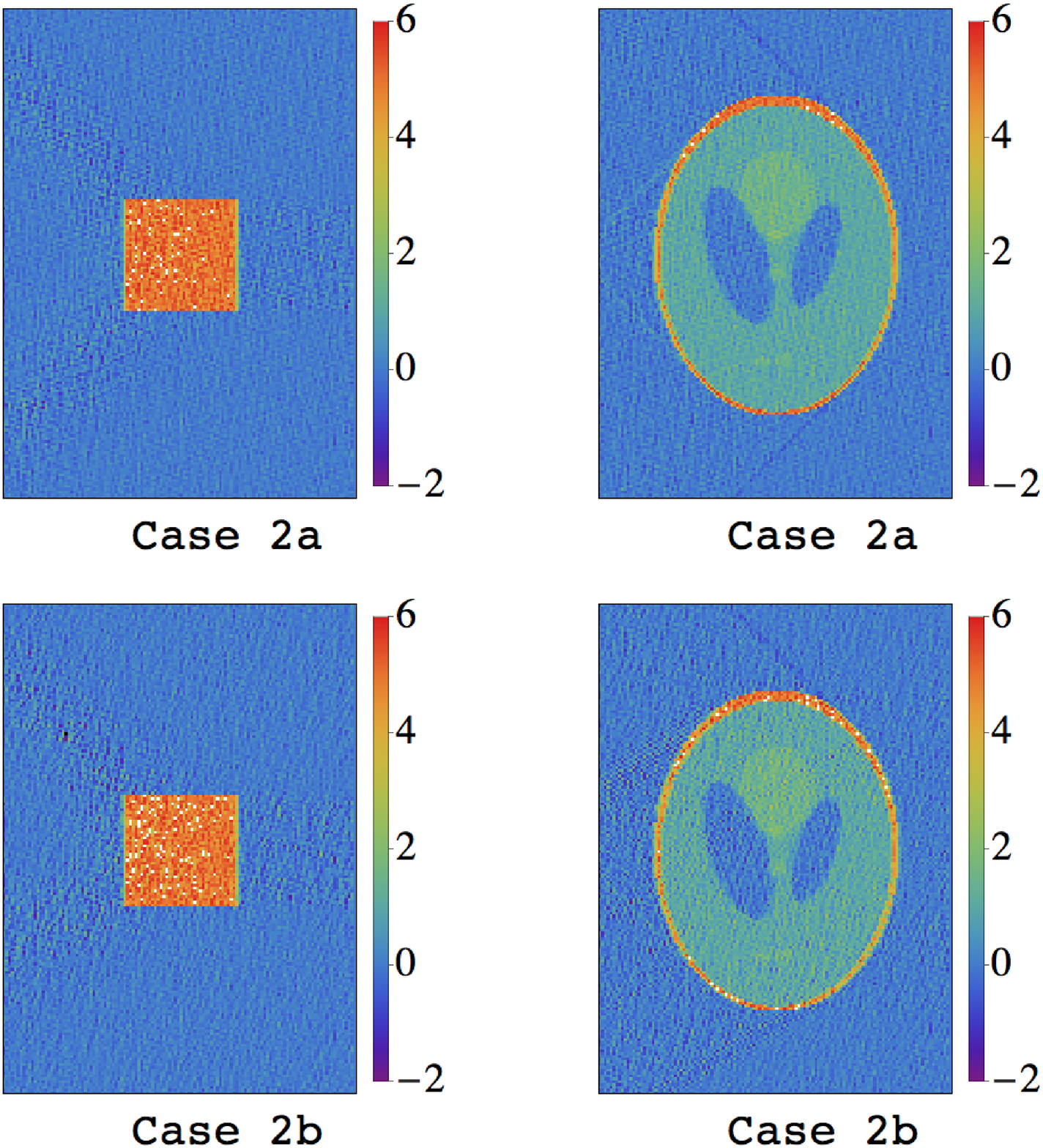,width=2.6in}}
}
\centerline{
\subfigure[${\mathcal N} = 10^4$, $\lambda= 0$]{
\epsfig{file=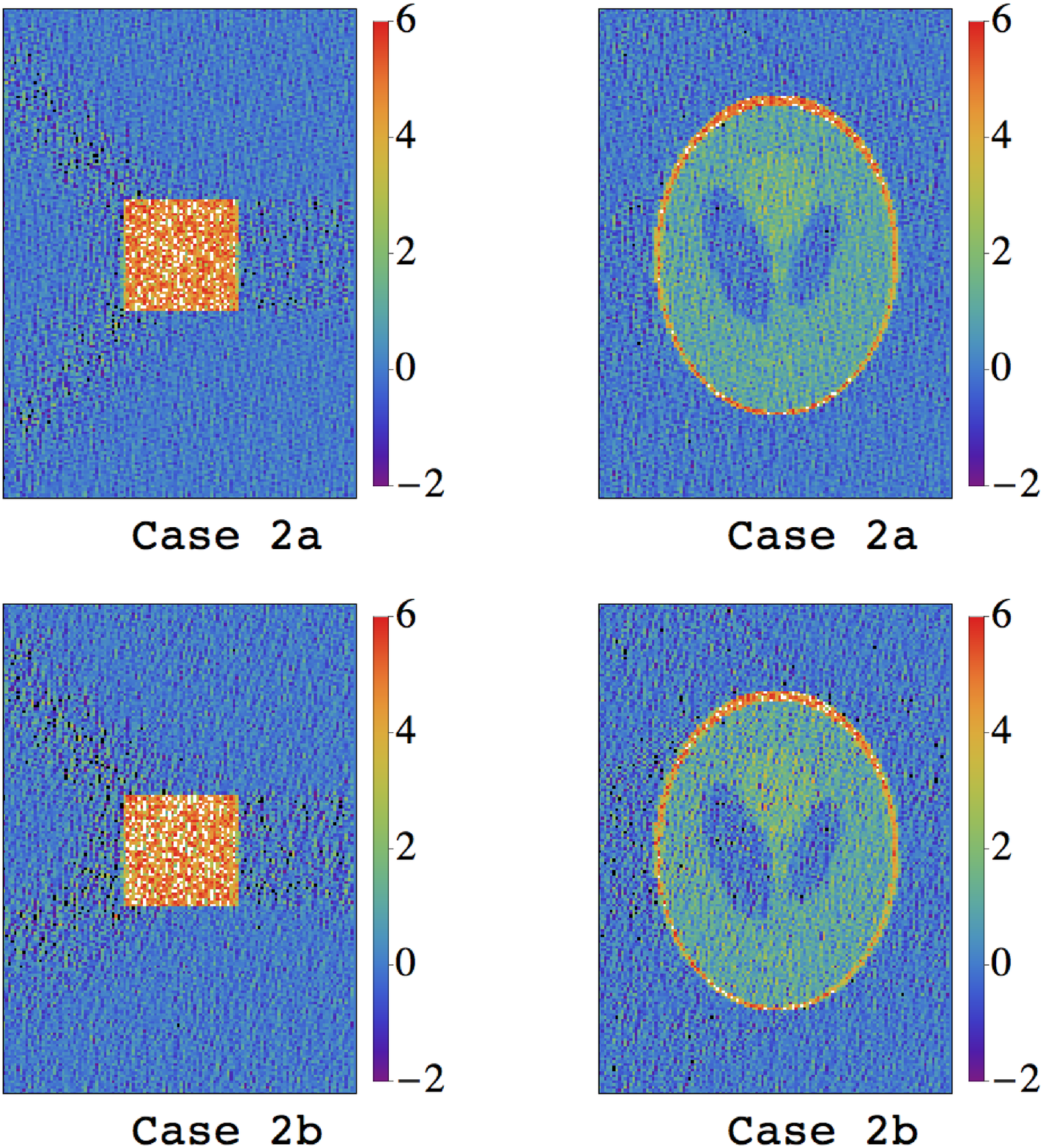,width=2.6in}}
\subfigure[${\mathcal N} = 2.5 \times 10^3$, $\lambda= 0$]{
\epsfig{file=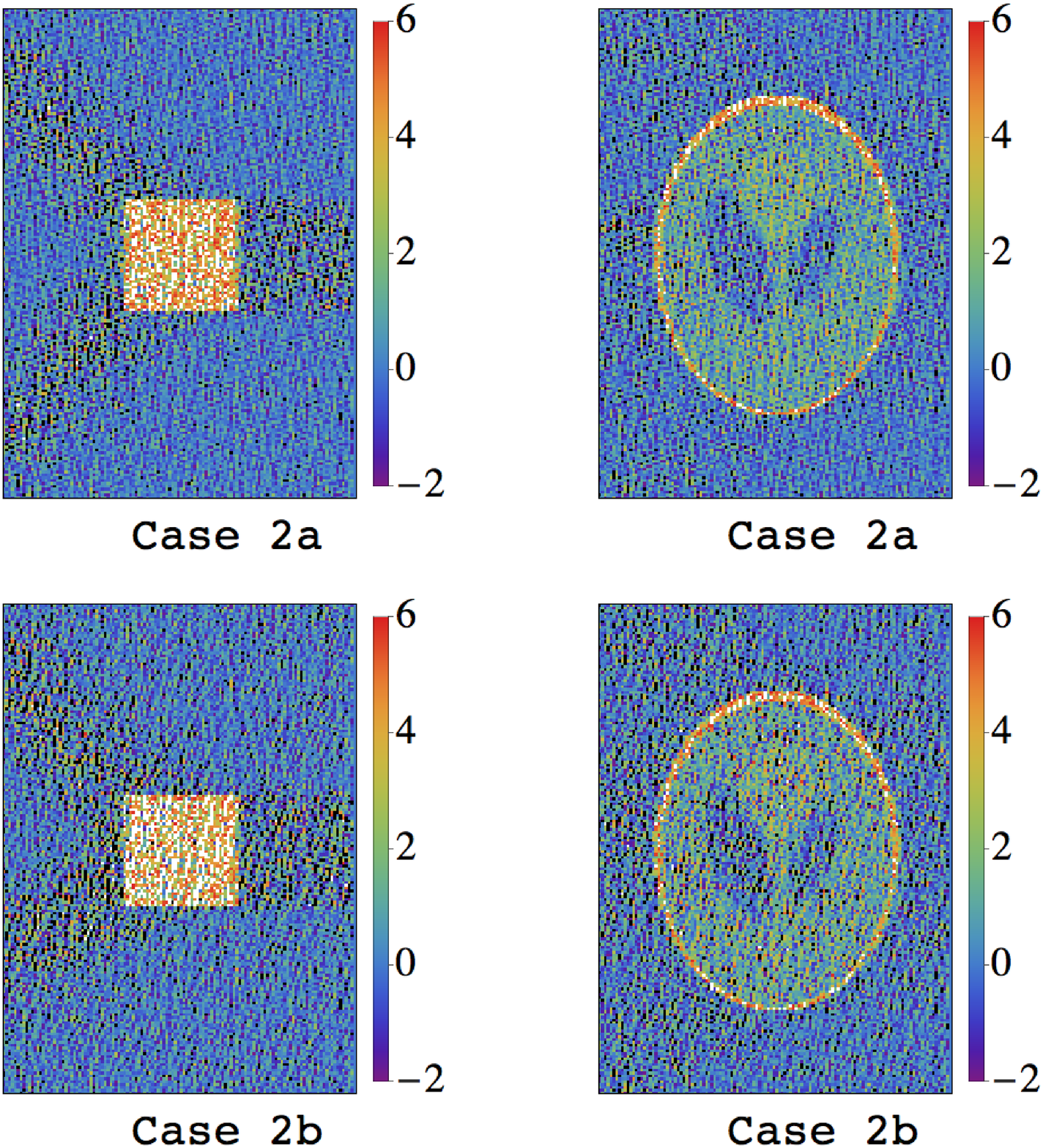,width=2.6in}}
}
\caption{\label{fig:Case_2} (color online) Same as in Fig.~\ref{fig:Case_1} but for
  Case 2; $\lambda=0$ in all cases (regularization is not used).}
\end{figure}

We now turn to Case 2 ($K=3$; see Table~\ref{tab:1} for more detail).
Similarly to Case 1, Case 2 is only applicable to reconstructing
absorptive contrast in a medium with a spatially-uniform scattering
coefficient. Reconstructions are shown in Fig.~\ref{fig:Case_2}. In
Case 2, the function $f(\theta)$ does not have zeros and conditioning
of the inverse problem is significantly improved. Correspondingly, we
have obtained reasonable reconstructions at all noise levels
considered without regularization, that is, by using $\lambda=0$.
Analytically, pseudo-inverse with $\lambda = 0$ is indistinguishable
from the ordinary inverse, and we have verified this fact numerically.

\begin{figure}
  \centerline{\epsfig{file=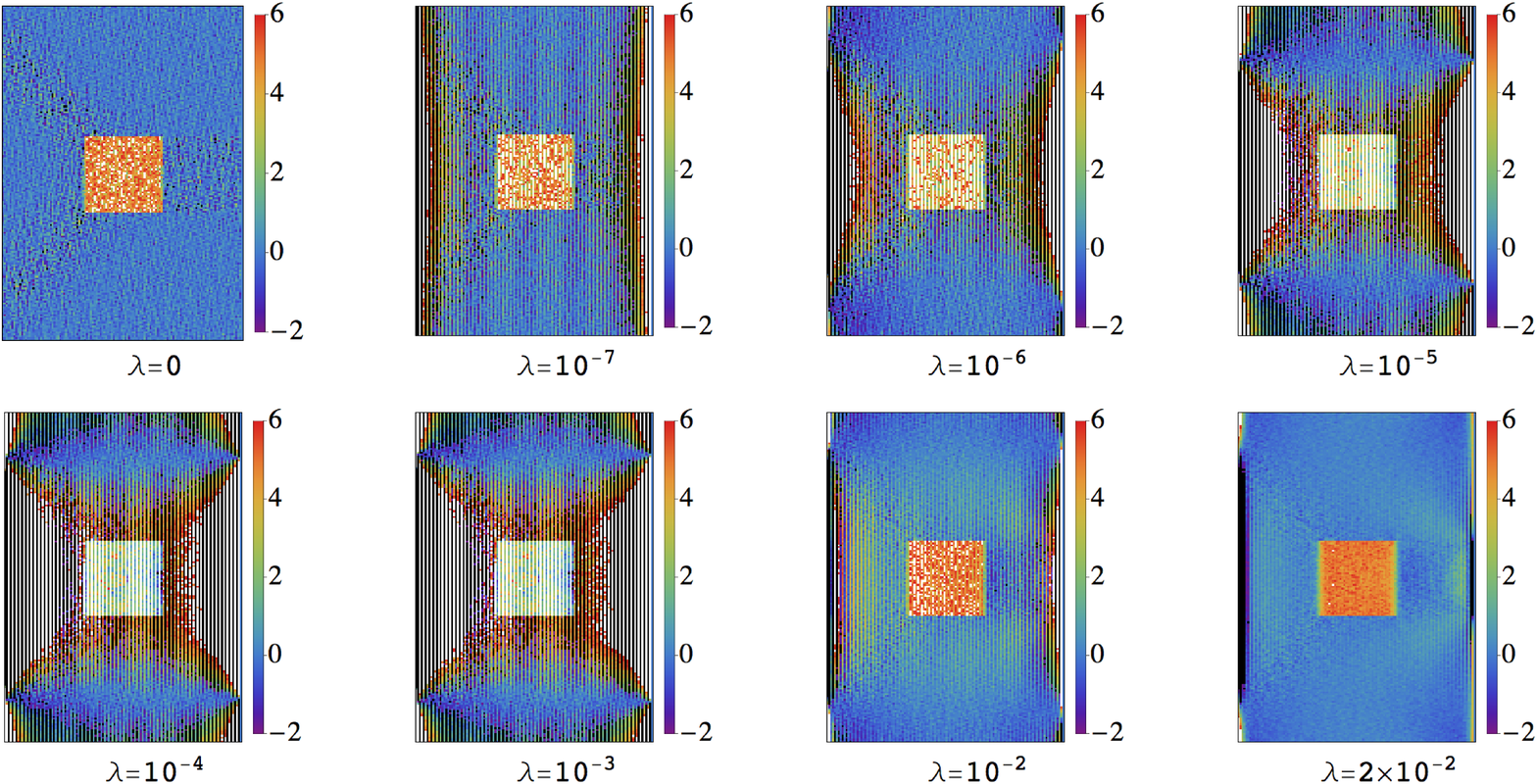,width=5.3in}}
\caption{\label{fig:Case_2_reg} (color online) Case 2a:
  Effects of regularization for the noise level ${\mathcal N} = 10^4$
  and the regularization parameter $\lambda$, as labeled.}
\end{figure}

Although Fig.~\ref{fig:Case_2} shows reasonable reconstructions
obtained with $\lambda=0$, the effects of regularization deserve
additional discussion. We have verified that in Case 2 regularization
does not yield noticeable improvement of image quality at all noise
levels used. This result is expected when one is inverting a
well-posed operator such as the forward operator of the discrete
Fourier transform, which has a flat spectrum of singular values. In
such cases, introduction of Tikhonov regularization is not justified.
However, in the case considered here, the spectrum of singular values
is not flat. Apparently, there exists at least one singular value,
which is numerically small, yet not small enough to cause significant
instability at $\lambda=0$. Under the circumstances, regularization
with an inappropriately chosen parameters $\lambda$ can produce image
artifacts, as is illustrated in Fig.~\ref{fig:Case_2_reg}. It can be
seen that the images with $\lambda=0$ and $\lambda=2 \times 10^{-2}$
are comparable in quality, although in the first case the
reconstructed boundary of the square inhomogeneity is sharper while in
the second case the image appears to be less noisy. At the
intermediate values of $\lambda$ (e.g., $\lambda = 10^{-4}$) a severe
artifact appears in the reconstructions. The artifact has the form of
oscillations whose wave vector is aligned with the $Z$-axis. We
attribute this to the fact that the Fourier series expansion of the
image does not converge in the usual sense at $q=0$, as was discussed
in Sec.~\ref{subsec:q=0} [after Eq.~(\ref{mu_n_q=0_sol})]. Note, however,
that in the more practically-important Case 3, this type of artifacts
is not present.

\subsection{Case 3}

\begin{figure}
\centerline{\subfigure[Case 3a]{\epsfig{file=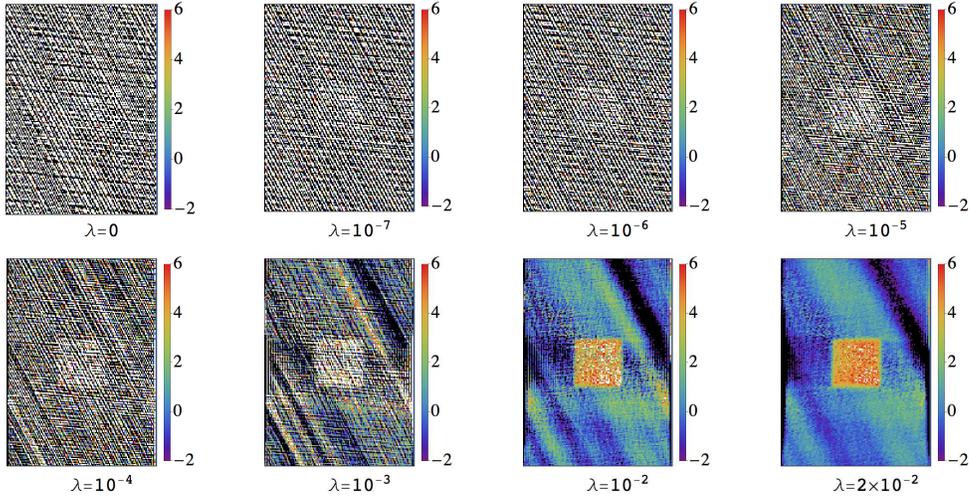,width=5.3in}}}
\centerline{\subfigure[Case 3b]{\epsfig{file=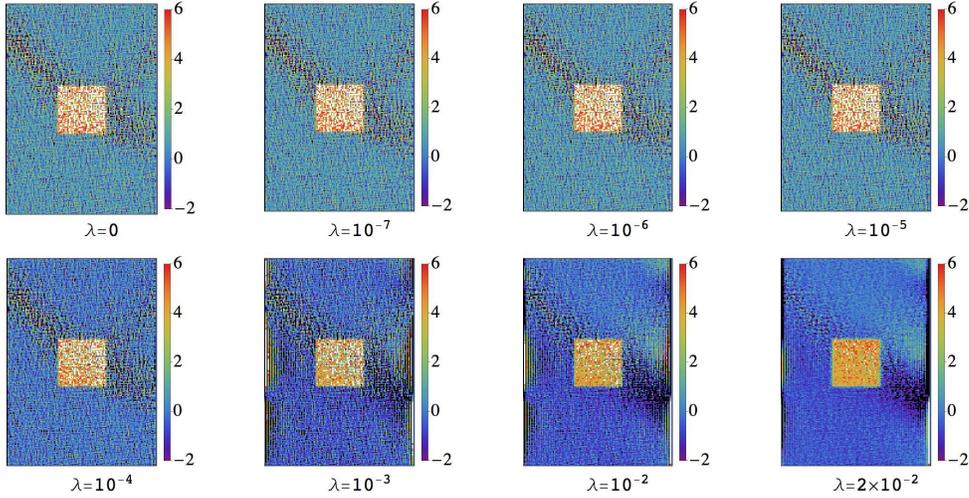,width=5.3in}}}
\caption{\label{fig:Case_3_reg} (color online) Case 3: Effects of regularization for
  the noise level ${\mathcal N} = 10^4$ and the regularization
  parameter $\lambda$, as labeled.}
\end{figure}

We now turn to Case 3 ($K=3$; see Tab.~\ref{tab:1} for more detail).
Here the weight coefficients $s_k$ satisfy the condition $\sum_k s_k =
0$, which makes possible simultaneous reconstruction of attenuation
and scattering. Note that the Cases 3a and 3b correspond to the cases
(a) and (b) illustrated in Figs.~\ref{fig:dirs} and \ref{fig:ftheta}.
The function $f(\theta)$ has zeros in Case 3a but not in Case 3b.
Correspondingly, Case 3b allows for a stable inversion. Indeed, it can
be seen that the reconstructions in Case 3b are much more stable in
the presence of noise than in Case 3a. Without regularization,
addition of noise to Case 3a results in noisy images that do not
resemble the phantom. This can be alleviated by using Tikhonov
regularization, as shown In Fig.~\ref{fig:Case_3_reg}. However, when
compared at the same level of regularization, the image quality is
always better in Case 3b. Note that the oscillating artifact that was
seen in Case 2 at intermediate values of $\lambda$
(Fig.~\ref{fig:Case_2_reg}) is not present in Case 3.

\begin{figure}
\centerline{
\subfigure[Case 3a]{
\epsfig{file=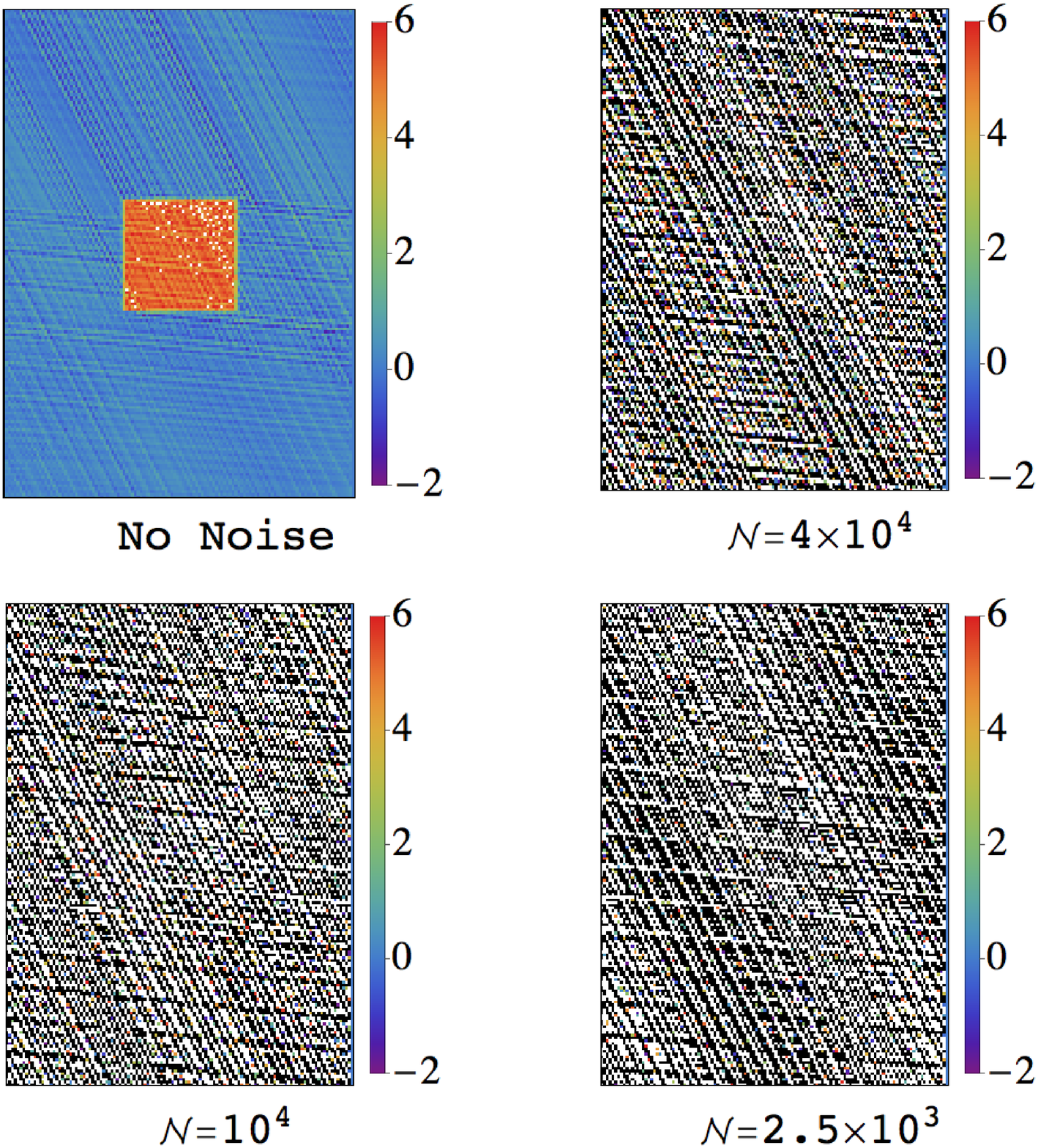,width=2.6in}}
\subfigure[Case 3b]{
\epsfig{file=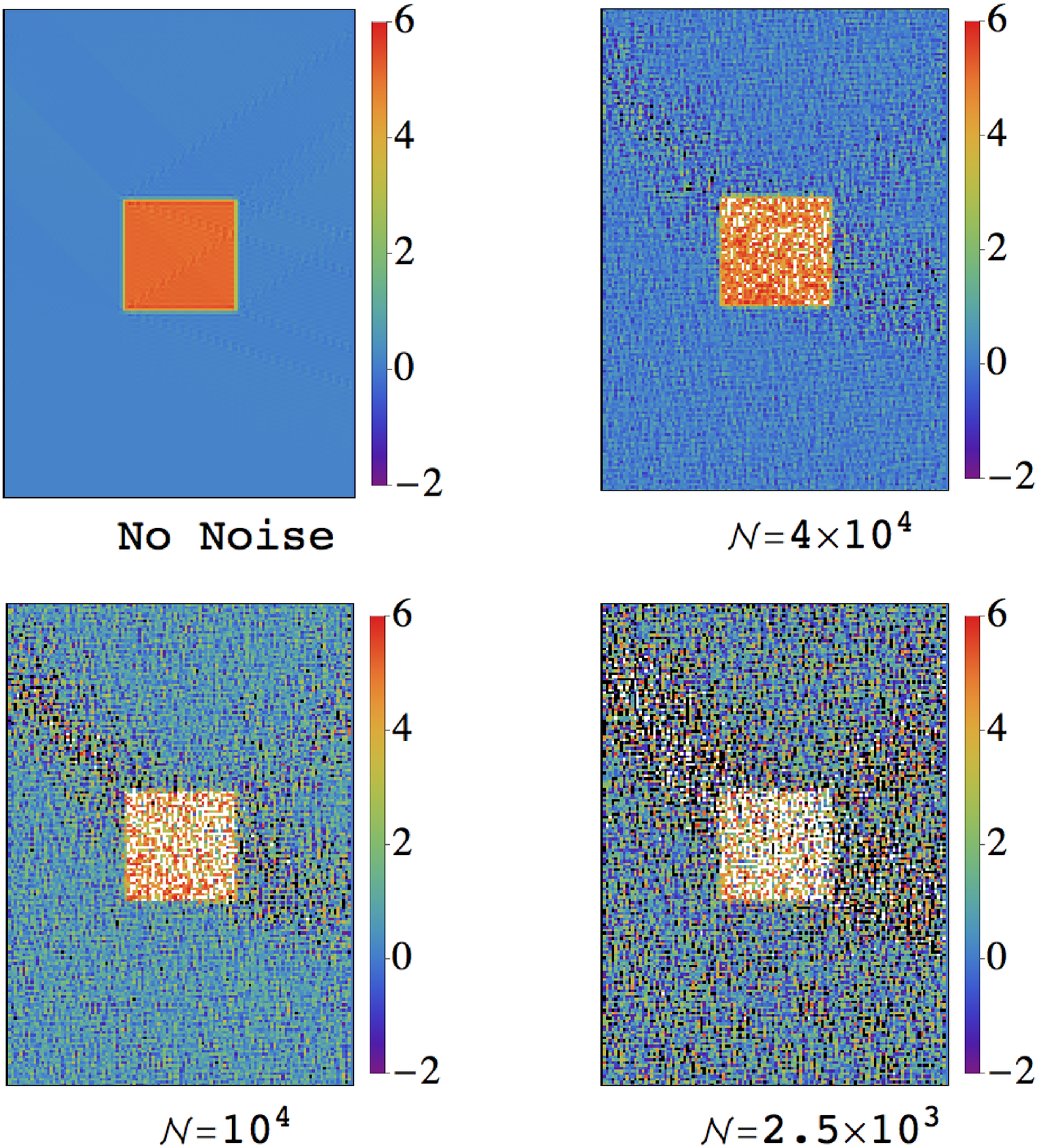,width=2.6in}}
}
\caption{\label{fig:Case_3} (color online) Case 3: Reconstructions of a square
  inhomogeneity for different levels of noise and $\lambda=0$ (without
  regularization).}
\end{figure}

\section{Discussion}
\label{sec:disc}

In this paper, we have introduced the star transform as a
generalization of the broken-ray transform, which was studied by us
previously in Refs.~\cite{florescu_09_1,florescu_10_1,florescu_11_1}.
We have shown that this generalization can be useful for improving the
stability of inversion. Theoretical considerations demonstrate that a
linear combination of individual measurements that leads to the star
transform is required when the scattering coefficient of the medium is
not spatially uniform and if one wishes to formulate the inverse
problem with respect to only one unknown function, $\mu({\bf R})$. The
previously-developed approaches to inverting the broken ray transform
in a medium with non-uniform
scattering~\cite{florescu_10_1,florescu_11_1} were, in fact, special
cases of the more general framework introduced in this paper. This
framework, based on constructing and inverting the star transform, can
be used to avoid ill-posedness and thereby improve the image quality.

Easy-to-use necessary conditions for avoiding the ill-posedness have
been formulated in Sec.~\ref{sec:stab}. However, these conditions are
not sufficient. Conditions for numericakl stability has been formulated 
separately in the limits $qL \ll 1$ and $qL \gg 1$.
A necessary and sufficient condition of numerical stability for $qL \ll 1$
is $\Sigma_0 \neq 0$, $\Sigma_1 \neq 0$, where $\Sigma_k$ are defined in 
(\ref{S_012_def}). However, if measurements of ballistic rays are used,
the condition $\Sigma_0\neq 0$ is no longer necessary.
In the limit $qL \gg 1$, a necessary and sufficient condition for stability 
can be obtained by plotting the function $f(\theta)$ [defined in (\ref{f_def})] and
visually determining whether it has zeros in the interval $0 \leq
\theta \leq \pi$. For intermediate values of $q$, no analytical results
concerning numerical stability have been obtained. However, we have seen numerical
evidence that $A(q)$ can be ill-posed for $qL\sim 1$ 
in the imaging geometries where all rays involved crossed the
same boundary.

We have also developed in Sec.~\ref{sec:methods}
several computationally efficient methods for inverting or
pseudo-inverting a matrix of the form (\ref{A_gen}). Fourier-space
inversion of the star transform is obtained as a special case.
However, the results of Sec.~\ref{sec:methods} may have a broader
utility because matrices of the form (\ref{A_gen}) are commonly
encountered in applications.  

Finally, we have provided an initial
numerical test of an imaging modality based on inverting the star
transform. We have reconstructed both a square phantom and the
Shepp-Logan phantom at various levels of Poissonian noise. Imaging
geometries suitable for simultaneous reconstruction of the scattering
and attenuation coefficients have been used, although in this paper
only reconstruction of attenuation has been demonstrated.

Therefore, we have shown that the star transform is a feasible
approach to imaging the medium with the account of single scattering.
But is formation of the star transform necessary?  As is described in
detail in Sec.~\ref{sec:derivation}, the star transform is obtained by
taking certain linear combinations of individual ``measurements''
$\phi_{jk}({\bf R})$. Of course, one can not claim that taking these
linear combinations provides additional useful information about the
image. In fact, the complete set of measurements $\phi_{jk}({\bf R})$
contains all experimentally-available information. The question is,
therefore, whether one should use the functions $\phi_{jk}({\bf R})$
directly or form the star transform.

The answer to this question depends on the numerical method used for
reconstruction, available computational resources and statistical
properties of the measurement noise. One possible approach is the
following. Let us take all $K(K-1)/2$ mathematically-independent
measurements $\phi_{jk}({\bf R})$ and discretize the equations
(\ref{phi_ij_mu}) on a square grid of the size $N \times N$. A
discretization scheme relevant to ray integrals is described, for
example, in~\cite{yuasa_97_1}, and we have used a similar
discretization approach in~\cite{florescu_09_1}. This will result in a
system of $K(K-1) N^2 / 2$ linear equations with respect to $2 N^2$
unknowns $\mu_{lm}$ and $\eta_{lm}$, $l,m=1,\ldots,N$. In practice, we
can (and should) assume {\em a priori} that the functions to be
reconstructed are zero in the pixels adjacent to the strip boundaries;
this will reduce somewhat the number of unknowns. This system of
equations can be viewed as a problem of optimization, which can be
solved by a variety of methods, including computation of the
pseudo-inverse, TV-regularization or iterative optimization with
nonlinear constraints. All these methods have considerable advantages
and should be investigated in their own right. However, an important
limiting factor is the computational complexity. Assume that we are
aiming at reconstructing a megapixel image, i.e., a rasterized 
$N \times N$ image with $N\sim 10^3$.  The computational complexity 
of the methods just described is
prohibitively high in this case even for reconstructing a single slice
of the medium. On the other hand, the methods based on constructing
and inverting the star transform are characterized by a much smaller
computational complexity because some of the steps necessary for image
reconstruction are performed analytically rather than numerically.

Let us estimate the computational complexity of solving the
image-reconstruction problem just considered. We assume that the
number of data points is of the same order as the number of unknowns
and will disregard factors of the order of unity. A direct method
(matrix inversion or computing the regularized pseudo-inverse)
requires $(N^2)^3 = N^6 \sim 10^{18}$ floating-point operations. This
is clearly out of reach, even with the use of supercomputers. The only
feasible option is in this case to use iterative optimization in which
the computational cost of each iteration scales as $(N^2)^2=N^4 \sim
10^{12}$. On an average modern computer with the peak performance of
$10{\rm Gflops}$, one iteration will cost about $100 {\rm sec}$ of
computational time. This is acceptable as long as only a few
iterations are required, but there is little hope that this would be
generally the case, especially if optimization with nonlinear
constraints is used. On the other hand, the iterative method of
Sec.~\ref{subsec:A_pseudo} require only $K N^3 \sim K \cdot 10^9$
floating-point operations per slice, where $K$ is of the order of
unity. This means that one slice can be reconstructed in less than a
second. The direct method of Sec.~\ref{subsec:dir} is even faster. The
practical improvement of utilizing the star transform is therefore
obvious.

In summary, construction and numerical inversion of the star transform
as described in this paper is a computationally efficient approach for
image reconstruction applicable to rays or particles undergoing
predominantly single scattering. In the case of X-ray imaging,
utilization of single-scattered photons is expected to reduce the
total radiation dose received by a patient. Applications of the
proposed methodology to mesoscopic optical imaging can also be
envisioned.

\section*{Acknowledgments}

This research was supported by the NSF under the grant DMS-1115616 to VAM and 
the grants DMS-1115574 and DMS-1108969 to JCS. The authors are grateful to
Guillaume Bal and Alexander Katsevich for valuable discussions.

\section*{References}

\bibliographystyle{iopart-num}
\bibliography{abbrev,master,book}

\end{document}